\documentstyle[11pt,aaspp4]{article}
\received{....}
\journalid{337}{....}
\articleid{11}{14}
\slugcomment{Accepted for publication on  ApJ}

\begin{document}
   \title{The REX survey: a search for Radio Emitting
X-ray sources.}
\author{Alessandro~Caccianiga\altaffilmark{1}, Tommaso~Maccacaro, 
Anna~Wolter, Roberto~Della Ceca,}

   \affil{Osservatorio Astronomico di Brera, Via Brera
28, I-20121 Milano, Italy}
\begin{center}
and
\end{center}
\author{Isabella~M.~Gioia\altaffilmark{2}}

   \affil{Institute for Astronomy, 2680 Woodlawn Drive, Honolulu, 
HI, 96822 USA}
\altaffiltext{1}{Present Address: Observatorio Astronomico de Lisboa, 
Tapada da Ajuda, P-1300 Lisboa, Portugal}
\altaffiltext{2}{also Istituto di Radio Astronomia del CNR, via Gobetti 101, 
40129 Bologna, Italy}
 
   
\begin{abstract}
We present the scientific goals, the strategy and the first results 
of the REX project, 
an effort aimed at creating a sizable and statistically complete sample of 
Radio Emitting X-ray sources (REX) using the available data from a VLA 
survey (NVSS) and the ROSAT PSPC archive.
Through a positional cross-correlation of the two data sets we
have derived a sample of about 1600 REX.
Among the 393 REX identified so far (either from literature or from our own  
spectroscopic observations) a high fraction is represented by AGNs 
(about 60 - 80\%), typically radio loud 
QSOs and BL Lacs.  
The remaining sources are galaxies, typically radio galaxies isolated or in cluster. 
Thanks to the low flux limits in the radio (5 mJy at 1.4 GHz) and in the 
X-ray band ($\sim 5\times 10^{-14}$ erg 
s$^{-1}$ cm$^{-2}$, 0.5--2.0 keV) and the large area of sky 
covered by the survey (2183 deg$^2$), we 
intend to derive a new complete and unbiased sample of BL Lacs which 
will contain both ``RBL'' and ``XBL'' type objects. In this way, the apparent 
dichotomy resulting from the current samples of BL Lacs will be directly 
analyzed in a unique sample. Moreover, the high number of BL Lacs expected in 
the REX sample ($\sim$ 200) will allow an accurate estimate of 
their statistical properties, like the X-ray, radio and optical 
luminosity functions and the cosmological evolution. 
For these reasons, the REX sample will be 
a powerful tool to test accurately the current theoretical models 
proposed for BL Lacs. To date, we have discovered 15 new BL Lacs and 
11 BL Lac candidates with  optical properties intermediate between those 
of a  typical elliptical galaxy and those of a typical  BL Lac object. 
These objects could harbour weak sources of non-thermal continuum in 
their nuclei and, if confirmed, they could represent the faint tail of the 
BL Lac population. The existence of such ``weak'' BL Lacs is matter of 
discussion in recent literature 
and could lead to a re-assessment of the defining criteria of a BL Lac 
and, consequently, to a revision of their cosmological and statistical 
properties. 

Finally the sample of $\sim$800 Emission Line AGNs, resulting from the 
REX survey, will be useful in addressing 
many of the open questions regarding the AGN phenomenology like the 
relationship between radio loud and radio quiet AGNs. 

\end{abstract}


%
%

\section{Introduction}

The optical identification of radio or X-ray catalogs is instrumental 
in deriving large samples of AGNs (e.g, Stocke et al. 1991, 
Boyle et al. 1994, Page et al. 1996 for the X-ray band
and Gregg et al. 1996 for the radio band).  
In these years, new radio and X-ray surveys characterized by low flux limits
and a wide coverage of the sky are in progress. In the radio band 
there are two similar projects aimed at creating new catalogs of 
radio sources at mJy fluxes with accurate radio positions (few arcseconds of
uncertainty or better)
based on VLA observations, i.e. the NRAO VLA Sky Survey (NVSS, Condon et al. 
1998) and the Faint Image of Radio Sky at Twenty centimeters survey (FIRST 
survey, Becker et al. 1995). The NVSS survey covers all the sky 
north of $\delta = -40^{\circ}$ using 
the VLA D and DnC configurations, with a typical rms of 0.45 mJy;
the FIRST survey covers about
10,000 $\deg^2$ around the North Galactic Pole plus a southern strip 
(from RA = 21$^h$ 20$^m$ to 3$^h$ 20$^m$ and from Dec = --2.5$^{\circ}$ to
1.6$^{\circ}$) with a 
higher spatial resolution (B configuration) and a lower flux limit 
(rms $\sim$ 0.15 mJy) with respect to the NVSS. Eventually, both catalogs will 
contain more than 10$^6$ radio sources each. In the X-ray band, 
the ROSAT satellite has produced a great bulk of data, both from the  
All Sky Survey (RASS) and from about 4000 PSPC pointed images gathered into 
public archives. The first catalog of bright (f$_{[0.1-2.4]keV}>$10$^{-12}$
erg s$^{-1}$ cm$^{-2}$) X-ray sources produced by the
RASS contains about 19,000 sources (Voges et al. 1996), while the catalogs 
derived on the
basis of public pointed PSPC images (WGA catalog, White, Giommi and  
Angelini, 1994 and ROSATSRC catalog, ROSAT NEWS n.32, 1994) contain 
about 70,000 sources each. 
All these surveys offer a great opportunity to derive new sizable 
samples of AGNs. In particular, the RASS and the VLA 
surveys will be useful to perform detailed statistical analyses. 
On the other hand, the large numbers of sources contained in these 
catalogs draw the attention to the problem of the optical 
identification.  Even with the high positional accuracy of the VLA, 
which
guarantees the (almost) univocal identification of the correct optical 
counterpart (at least for $m_V$ $\leq$ 20), the number of 
optical candidates to observe spectroscopically
is undoubtedly high. This number increases dramatically in the case of the 
X-ray catalogs derived from the ROSAT PSPC data, which are characterized by 
larger positional errors 
(typically from 14$\arcsec$ to 50$\arcsec$, see \S~4.3). 
As a consequence, the completion of the optical   
identification process of these surveys will require decades. 
This fact points out the importance of finding efficient  
pre-selection techniques to extract from the whole sample only
the candidates of
interest for spectroscopic follow up. Many authors have 
initiated 
specific efforts aimed at creating samples of AGNs from VLA data 
(Gregg et al. 1996) or BL Lacs from ROSAT data
(Perlman et al. 1998; Laurent-Muehleisen et al. 1998; Nass et al. 1996;
Kock et al. 1996) through the application of particular pre-selection 
criteria based on 
radio-to-optical or X-ray-to-optical flux ratios, or using information 
on optical polarization.

The primary goal of the REX project is
the selection of a {\it statistically complete} sample of AGNs,
in particular radio loud  (RL) QSOs and BL Lacs, by using {\it simultaneously} 
the  
information derived from the NVSS survey and from pointed ROSAT PSPC 
images. Through a positional 
cross-correlation of sources detected in the NVSS and ROSAT fields, 
we derive a sample of {\it Radio 
Emitting X-ray sources (REXs)} which is expected to contain a high fraction 
of AGNs (radio loud QSOs, BL Lacs, Seyfert galaxies) and radio 
galaxies. This method has two fundamental advantages: 

1) We are able to minimize ``a priori'' the presence in the sample of 
objects like stars or ``normal'' 
galaxies, which are not strong radio emitting X-ray sources, increasing 
 in this way the efficiency of finding the sources of interest (BL Lacs, QSOs) 
to be observed spectroscopically;

2) We can use the accurate VLA positions to pinpoint the optical counterparts
of the X-ray sources.

In this paper, we discuss the main scientific goals of 
the project (\S~2) and the catalogs of radio (\S~3) and 
X-ray sources 
(\S~4) used for the cross-correlation. In \S~5 we discuss  
the source selection 
criteria and the estimated content of the REX survey. The strategy of 
optical identification of the sample is described in \S~6 while 
the discussion of the content of the REX sample is reported in \S~7. 
In \S~8 we report preliminary results derived from the study of the 
sample of 
BL Lacs discovered so far. Our conclusions are summarized in \S~ 9. 

Throughout the paper, the values of $H_0$ = 50 Km s$^{-1}$ Mpc$^{-1}$ and 
$q_0$ = 0 are used.

\section{Scientific goals}
The main goal of the project is the study of RL AGNs and
 BL Lacs 
objects. These kinds of AGN are either intrinsically few (RL AGNs are about 
10\% of the whole class of AGN
\footnote{This is true in the optically selected samples, while in 
the X-ray selected samples, e.g. in the EMSS sample, the fraction of RL AGNs 
ranges from 3\% to 30\%, depending on the optical luminosity (e.g. Della Ceca 
et al. 1994). We consider radio loud an AGN with $\alpha_{ro} > 0.35$ 
(see Della Ceca et al. 1994), where $\alpha_{ro} = log (S_{5 GHz}/S_{2500 A})/5.38$.} 
while BL Lacs represent only few percent) 
and/or hard to find. BL Lacs, in particular, lack 
a UV excess or any other peculiar optical spectral signature and, for these 
reasons, they 
are found with difficulty in the optical band. Nevertheless, they 
are both radio loud and X-ray-loud (e.g. Stocke et al. 1990) and, as a 
consequence, they are typically selected 
in the X-ray or in the radio band. The small numbers of BL Lacs or 
RL AGNs available in current complete samples (e.g. the EMSS catalog
 contains only 36 BL Lacs and 43 RL AGNs, Morris et al. 1991, Della Ceca et 
al. 1994  and the 1 Jy catalog contains only 34 BL Lacs, Stickel et al. 1991)
limits any detailed statistical analysis of these objects. 
On the other hand, accurate statistical studies are instrumental in order 
to address the 
 many open questions, like the origin of the differences between radio loud 
and radio  quiet  AGNs (e.g. Wilson \& Colbert 1995), the supposed dichotomy of 
 the BL Lac population (e.g. Giommi \& Padovani 1994, Padovani \& Giommi 1995), 
the cosmological 
properties (evolution, luminosity functions) of BL Lacs and the 
problem of their parent population (e.g. Urry \& Padovani 1995 and references 
therein). Many of these questions, in 
particular those related to BL Lacs, require a significant enlargement 
of the available complete samples. Moreover, 
many crucial predictions of the current theoretical models, like 
the relative number of radio selected (RBL) and X-ray selected (XBL) types (Giommi \& Padovani 1994, Fossati et al. 1997), could be verified 
only by reaching low X-ray flux limits 
(typically a few 10$^{-14}$ erg s$^{-1}$ cm$^{-2}$). The REX sample will 
possess these requirements, as it will contain $\gtrsim$ 200  BL Lacs 
down to a flux limit of $\sim 5 \times 10^{-14}$ erg s$^{-1}$ cm$^{-2}$ 
in the 0.5-2.0 keV energy band.

\section{The radio catalog}
The NRAO VLA Sky Survey  began in 1993 and  
covers now the whole sky north of $\delta=-40^{\circ}$ 
($\sim 10.3$ sr, 82\% of the total sky). It is performed at 1.4 GHz 
using the compact D and DnC configurations of the Very Large Array in Socorro
(New Mexico, USA). The primary product of this survey is a set of 2326 
$4^{\circ}\times4^{\circ}$ maps containing information about total-intensity and 
linear polarization (Stokes parameters I, Q and U). The size of the restoring beam 
($\theta =  
 45\arcsec$ FWHM) allows us to achieve a high surface-brightness
 sensitivity needed for
completeness and photometric accuracy. The rms of the Stokes I maps is 
$\sigma \sim0.45$ mJy beam$^{-1}$. The final maps are 
released electronically by anonymous FTP. Moreover, NRAO extracts a catalog 
of discrete sources and components (if the radio source is resolved by the 
VLA beam) detected in the images, by fitting elliptical 
Gaussian to all significant peaks. The entries consist of a single Gaussian
model fits. The sources (or components) are identified
in the Stokes I images and then the associated polarization information 
is derived from the Stokes Q and U images. A brief description of this 
catalog is given at the WWW home page of NVSS 
(http://www.cv.nrao.edu/$\sim$jcondon/nvss.html). 
A related program (NVSSlist) is also provided in order
to read the catalog (which is released in FITS format) and to browse 
selected portions of the  sky. NVSSlist corrects also the 
raw catalog for known biases and computes errors associated to the source 
model parameters (position, flux density, etc.). Both catalog and program 
are continually updated. For what concerns the REX project, we set the radio 
limiting flux to 5 mJy (about 10 times the rms of the I maps) in order to
achieve accurate positions (95\% error circles smaller than $5\arcsec$) 
and to guarantee the completeness of the sample (see Condon et al 1998 for 
details). We have used  version 2.8 of the NVSSlist program and 
the version 28 (01/19/98) of the catalog. In that version the NVSS 
was plagued by a number of holes, i.e. regions of missing data. 
We have estimated that in the region of sky  covered by the X-ray catalog
(see section~4) the area lost due to the holes is about 5-7\%. 

\section{The X-ray catalog}
The X-ray catalog used for the REX project has been derived from the ROSAT 
archive of pointed PSPC images. To date, the ROSAT PSPC archive represents
the best set of data to derive a large catalog of X-ray 
sources covering a wide angle of the sky  
and reaching faint limiting fluxes 
(below $10^{-14}$ erg s$^{-1}$ cm$^{-2}$ in the 0.2 - 2.4 kev band with the 
deepest images)\footnote{The RASS, although it covers the whole sky, reaches limiting 
fluxes of about $10^{-12}$ erg s$^{-1}$ cm$^{-2}$ (the Bright RASS
catalog) or fluxes of about 1--5$\times$10$^{-13}$ erg s$^{-1}$ cm$^{-2}$ 
(the total catalog of 60,000 sources presented in Voges, 1993) 
and, for this reason, 
it is not optimized to study the faint tail of the LogN-LogS of 
BL Lacs and RL AGNs.}. We briefly describe  
the PSPC archive and  the existing X-ray catalogs 
derived from these data. Since these catalogs are not suitable for 
our purposes, we have undertaken to reanalyze all the PSPC images 
suitable for this project with the aim of constructing a new statistically 
representative catalog of sources. We present the selection of the 
fields used in this 
project, the process of source detection and the source positional 
uncertainties in the resulting X-ray catalog. 

To date, all the pointed ROSAT PSPC observations ($\sim$4000 in total) are of public 
domain and they are stored in a public archive. 
The large field of view of the PSPC (about 2.8 deg$^2$) makes this archive 
of images particularly 
useful to build up an X-ray catalog of serendipitous sources. 
Two different catalogs of X-ray sources detected in a number of ROSAT PSPC 
images have been made public in the past years: the WGA catalog (White, Giommi and 
Angelini, 1994, IAU Circular 6100) and the ROSATSRC catalog (ROSAT NEWS 
n.32, 1994). 
The detection technique used for the WGA catalog is based on a
sliding box algorithm which is optimized for point-like sources. 
The inner (0$\arcmin$ -- 19$\arcmin$ off-axis angle)
and outer (18$\arcmin$  -- 55$\arcmin$ off-axis angle) regions of 
the images were processed separately to maximize the sensitivity to source 
detection. The latest version (rev. 1) of the catalog contains about 
68,000 sources detected in 3007 fields. 
The ROSATSRC catalog is the First Source catalog of pointed 
observations 
created at Max Planck Institut f\"ur Extraterrestrische Physik. A maximum 
likelihood  detection algorithm was used for the central 20$\arcmin$ of the 
PSPC field and a sliding box algorithm of fixed width was used at larger 
off-axis angles. 
About 74,000 sources are contained in the latest version of the catalog
(rev. 1) which is derived from the analysis of 3348 fields. 
Although the bulk of the images used are the same for both catalogs, 
there are significant differences, probably related to the different methods 
and thresholds used for the source detection (see White, Giommi and 
Angelini 1995, in the WGACAT WWW home page 
(http://lheawww.gsfc.nasa.gov/users/white/wgacat/wgacat.html)
 for a brief description of 
these differences). 
Both WGACAT and ROSATSRC catalogs were created principally to quickly derive 
 a list of X-ray sources useful, for instance, to find peculiar and 
interesting objects (e.g. sources with anomalous temporal variability or 
spectral properties) to be re-observed, if necessary, with ROSAT or other 
telescopes (for example ASCA). For this reason, they were created without a 
particular attention to the problem of completeness and representativeness.
Both catalogs contain, for example, several
repetitions, i.e. the same source may appear more than once in the two 
catalogs: for instance, about 4000 sources in the last version of WGACAT 
are redundant.
This problem is related to the fact that many 
PSPC pointings overlap one to another and a source 
may thus be detected separately in two different images. 
Moreover, the algorithms chosen for the source detection work on a fixed 
scale while the PSF of the PSPC changes considerably and continuously 
with the off-axis angle; this problem still holds even when using a 
different scale in the inner and in the outer region of the PSPC. 
All these problems make the definition of a complete sample and 
the derivation of the sky-coverage very difficult for both catalogs. 
Finally, WGA and ROSATSRC catalogs
 are not purely made of serendipitous sources because they contain 
all the targets of the PSPC pointings, affecting the representativeness 
of the two catalogs.  
For these reasons, WGACAT and ROSATSRC, although useful for
some specific purposes, are not suitable for statistical studies and, 
in particular, for our project, which requires   
completeness and representativeness besides a precise estimate of the 
sky-coverage.

Thus, we have reanalyzed a well defined subset (see \S~ 4.1) of the
public PSPC observations with the purpose of constructing
 a complete and well defined sample of X-ray sources. 
In particular, we pay attention: 

1) to resolve the problem of overlapping PSPC images;

2) to eliminate all the targets and the target-related sources from the 
catalog;

3) to use a detection  algorithm sensitive to the changing of the PSF over the 
   PSPC field;

4) to compute the appropriate sky coverage.

\subsection{Selection of ROSAT PSPC fields}
From the $\sim$ 4000 PSPC images  in the public archive,  
we have selected all the fields that fulfill the following criteria:

1) $\delta \geq -40^{\circ}$ (to match the NVSS sky coverage);

2) $|b^{II}| \geq 10^{\circ}$ (to avoid the galactic plane);

3) exposure time $\geq 1000 s$;

4) processed with the revision 7 of the SASS as of December 1997 (to 
use data with an improved aspect solution).

Moreover, we have excluded intrinsically ``confused'' regions 
like those with a high stellar-density (e.g. images pointed at
Pleiades, Hyades) or
near extended dark or molecular clouds, which could create 
problems in the process of source optical identification  or 
in  the computation of the limiting sensitivity of the field. 

\begin{figure}
\plotone{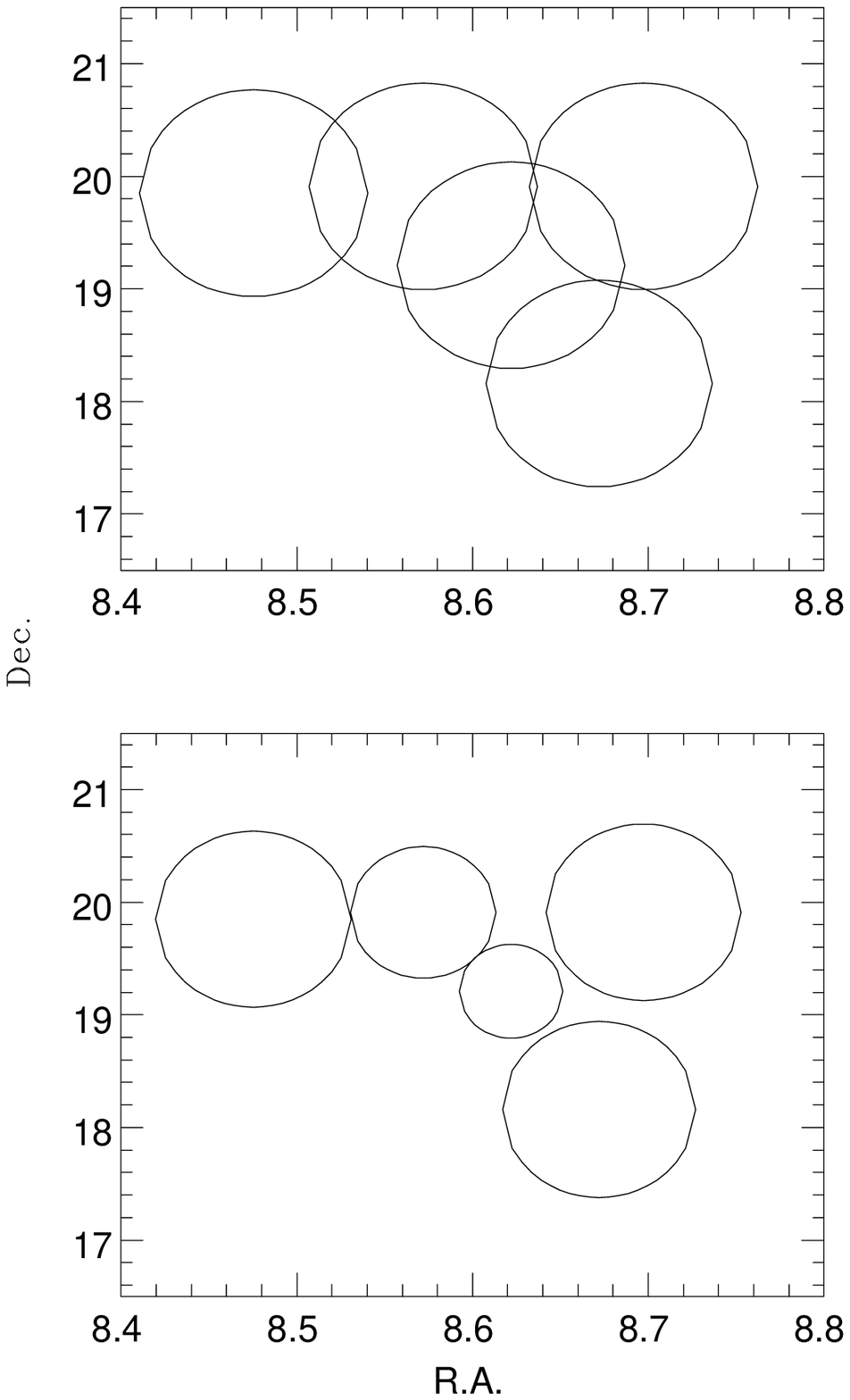}
\caption{(a) An example of overlap of PSPC fields. 
(b) The same region 
of sky after the cleaning procedure that produces a set of completely 
disjointed fields albeit of reduced size.
Although this procedure does not maximize the area
of sky covered (the region among the fields is not 
completely covered) it  simplifies the source detection procedure
that is applied on a set of totally independent fields. We also recall that a maximum 
radius of 47$\arcmin$ is used for the useful area of a PSPC image. 
\label{ovl}}
\end{figure}

The distribution in the sky of all the PSPC pointings 
clearly shows that 
there are several overlapping images (deep surveys, 
targets which cover a wide area of sky, fields 
pointed to close targets). Combining these images in single frames 
would create several problems during the process of source detection.  
Typically, the presence of strong gradients 
in the sensitivity profile of the final image, induced by the edges of the 
individual fields, creates difficulties to the background determination and 
to the detection algorithm and 
may yield spurious sources. We avoided this problem simply by 
excluding from the analysis the outer region of one 
of the overlapping fields in order to 
obtain a mosaic of separated fields. Figure~\ref{ovl} illustrates this 
``cleaning'' procedure. 

Finally, in the case of exact overlap in sky coordinates
of two or more distinct pointings, we 
summed them and we considered these fields as a 
single image with an exposure time which is the sum of the individual 
exposures. In this case, the name of the resulting field is formed by 
the lowest ROSAT Observation Request (ROR) sequence number with the suffix ``sum''.

In order to achieve a catalog of purely serendipitous sources, 
we have eliminated from the survey the area containing the target 
of the observation (since it is, by definition, not chosen randomly) 
and all the target-related sources. 
In the majority of the PSPC pointings, the 
target falls in the center of the field and its extent is typically 
included in a circle of 4$\arcmin$ radius. In the case of images 
pointed toward rich/nearby clusters of galaxies, the radius of the circle
around the target can be considerably larger (from 10$\arcmin$ 
to 20$\arcmin$). The few fields which contain very extended targets  
(dimension $\geq 25\arcmin$) have been fully excluded from the survey. 
In the few cases in which the target object 
was deliberately off-set from the center (e.g. for variability studies) 
we excluded a circle of radius $r_t$ centered on the target position. 
Furthermore, in the case of PSPC images centered on an optically
very bright object (e.g. $\alpha$ Ari, $\beta$ CMa), we have excluded the 
area ``blinded'' 
by the optical image. In these cases, in fact, it would be extremely 
difficult, 
if not impossible, to optically identify a serendipitous source close 
to the target. 

In total, we have selected 1202 distinct PSPC fields.
For reference, we have made electronically available\footnote{
The table pspc\_table.txt can be retrieved by ``anonymous'' ftp 
at {\it ftp.brera.mi.astro.it}.  
The table  (in ASCII format) is in the directory 
``/pub/REX/tables/''. A description of the table is found in 
the file pspc\_table.doc in the same directory.} 
a table containing the complete list of PSPC images used for this 
survey covering a total area of 2183 deg$^2$. In particular, we 
have reported, for each PSPC field, 
the Rosat Observation Request sequence number (ROR),
the sky coordinates (J2000), the exposure time (which is the total 
exposure time in the cases of summed fields), the 
inner ($r_i$) and the outer ($r_o$) radius of the PSPC region used, 
the position of the target in detector coordinates and the 
radius of the excluded area ($X_t$, $Y_t$ and $r_t$) if the 
target is off-set and the Galactic hydrogen column density 
(in units of 10$^{20}$ cm$^{-2}$) derived from Dickey \& Lockman (1990). 
When there were no targets (e.g. surveys or mispointed fields) 
both $r_i$ and $r_t$ are zero. 
The maximum value of the outer radius has been set to 47$\arcmin$ 
(although the ``standard'' radius of a PSPC field is 55$\arcmin$) 
in order to exclude the part of the PSPC 
where the PSF degradation is severe and the positional uncertainties
are too large ($>$ 50$\arcsec$).

\begin{figure}
\plotone{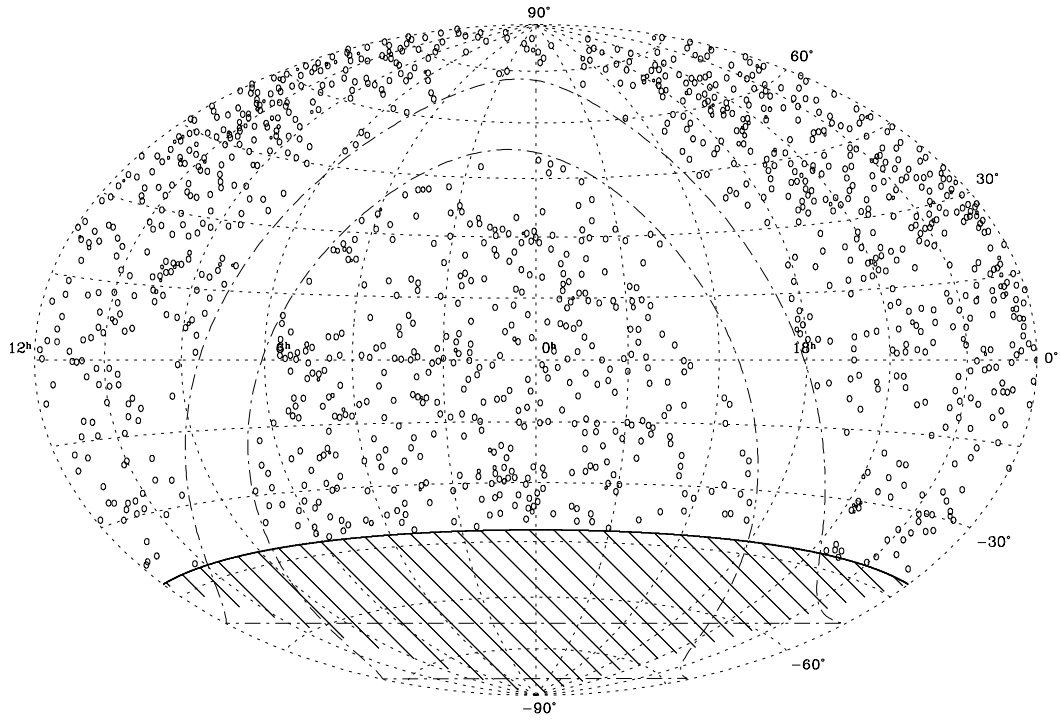}
\caption{Sky distribution of the PSPC fields used  
for the REX project. \label{ror_map}}
\end{figure}

\begin{figure}
\plotone{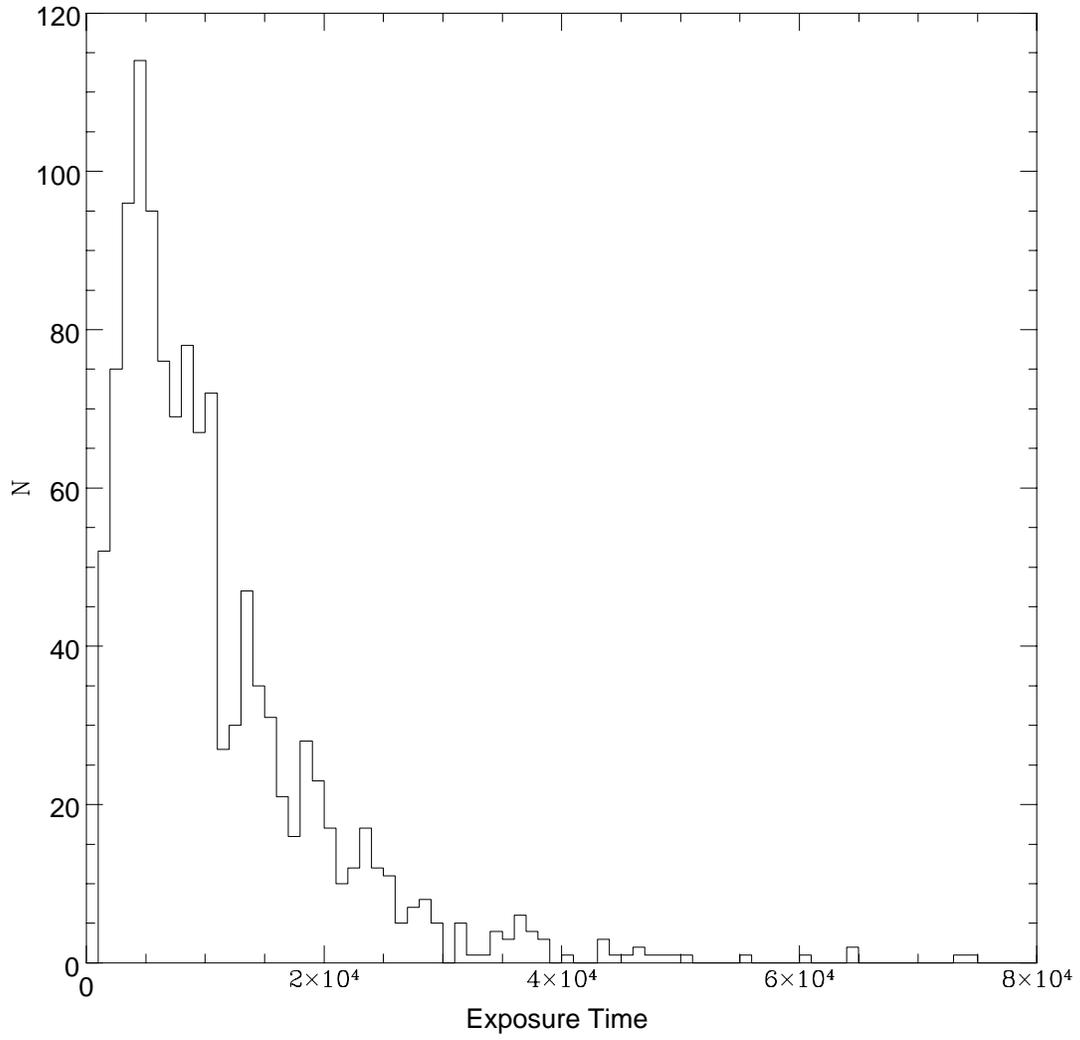}
\caption{The histogram of the exposure times of the used PSPC fields}
\label{ror_exp}
\end{figure}

We report in Figure~\ref{ror_map} the distribution in the sky of the total set
of selected PSPC fields while in Figure~\ref{ror_exp} we show the histogram of 
their exposure times.

\subsection{The detection algorithm} 
The detection algorithm used for this project was developed at the 
Palermo Astronomical Observatory and it is described in detail 
in Damiani et al. 1997a. The method  is based on wavelet transforms, 
that are 
well suited to the case where the Point-Spread Function (PSF) is
varying across the image (see also Rosati et al. 1995). 
This detection technique outperforms that used to produce the 
WGA and the ROSATSRC catalogs in terms of reliability and 
efficiency.

For the REX project, we have applied the detection algorithm to the 
selected set of PSPC fields, using the ``hard'' images (0.5 - 2.0 keV)
to reduce the intensity of the background and to minimize the effects 
due to Galactic absorption. Moreover, the choice of the hard band
minimizes the
uncertainties related to the conversion from count-rate to flux 
due to the unknown spectral shape of the sources, as discussed in 
Vikhlinin et al. (1995). The images used have 15$^{''}$ pixel size. 
The final catalog contains 16,275 X-ray 
sources detected with a confidence level $\geq$ 6$\sigma$ (see Damiani 
et al. 1997a for a discussion on the determination of the probability 
associated to a given source detection threshold) and a count-rate 
$\geq 3\times 10^{-3}$ cts/s. The high threshold on significance 
guarantees a very low number of spurious X-ray sources (see Figure 3 of 
Damiani et al. 1997b); the limit on the count-rate was chosen to avoid very
faint sources that
could be very difficult to identify spectroscopically.
Assuming a power-law spectrum with
an energy index  $\alpha_X = 1.0$ ($f_{\nu} \propto \nu^{-\alpha}$)
this count-rate limit corresponds to a limiting flux in the 0.5-2.0 keV band
of $\sim 3.5\times10^{-14}$ erg s$^{-1}$ cm$^{-2}$, for 
$N_H$ = 7$\times$10$^{19}$ cm$^{-2}$ (the lowest value for the 
PSPC fields used in the REX survey) and of 
$\sim$ 6$\times$10$^{-14}$ erg s$^{-1}$ cm$^{-2}$ 
for $N_H$ = 2$\times$10$^{21}$ cm$^{-2}$ (the highest value).

\subsection{The positional accuracy of the X-ray catalog}
Before cross-correlating the X-ray catalog with
that derived from the NVSS survey, it is crucial to  accurately
assess the positional uncertainties. 
The error circle of the sources detected 
in the ROSAT PSPC images is typically of the order of tens of
 arcseconds while the error circles associated with the NVSS positions are 
an order of magnitude smaller (few arcseconds). Therefore, the 
impact parameter, which is the square combination of the two error circles,
depends mainly on the X-ray positional accuracy. 

To determine the uncertainties associated with the X-ray positions, we have 
cross-correlated the X-ray sources against two catalogs:
the Hipparcos Input catalog of stars (HIC, Perryman et al. 1997)  and
the V\&V96 catalog  of AGNs (V\'eron-Cetty \& V\'eron 1996).

The HIC catalog contains 118000 stars with
a positional accuracy of $\sim$0.3$\arcsec$.  
In the case of the V\&V96 catalog we have used only the 8272 sources
with a good estimate of the position (accuracy
better than 1$\arcsec$).

\begin{figure}
\plotone{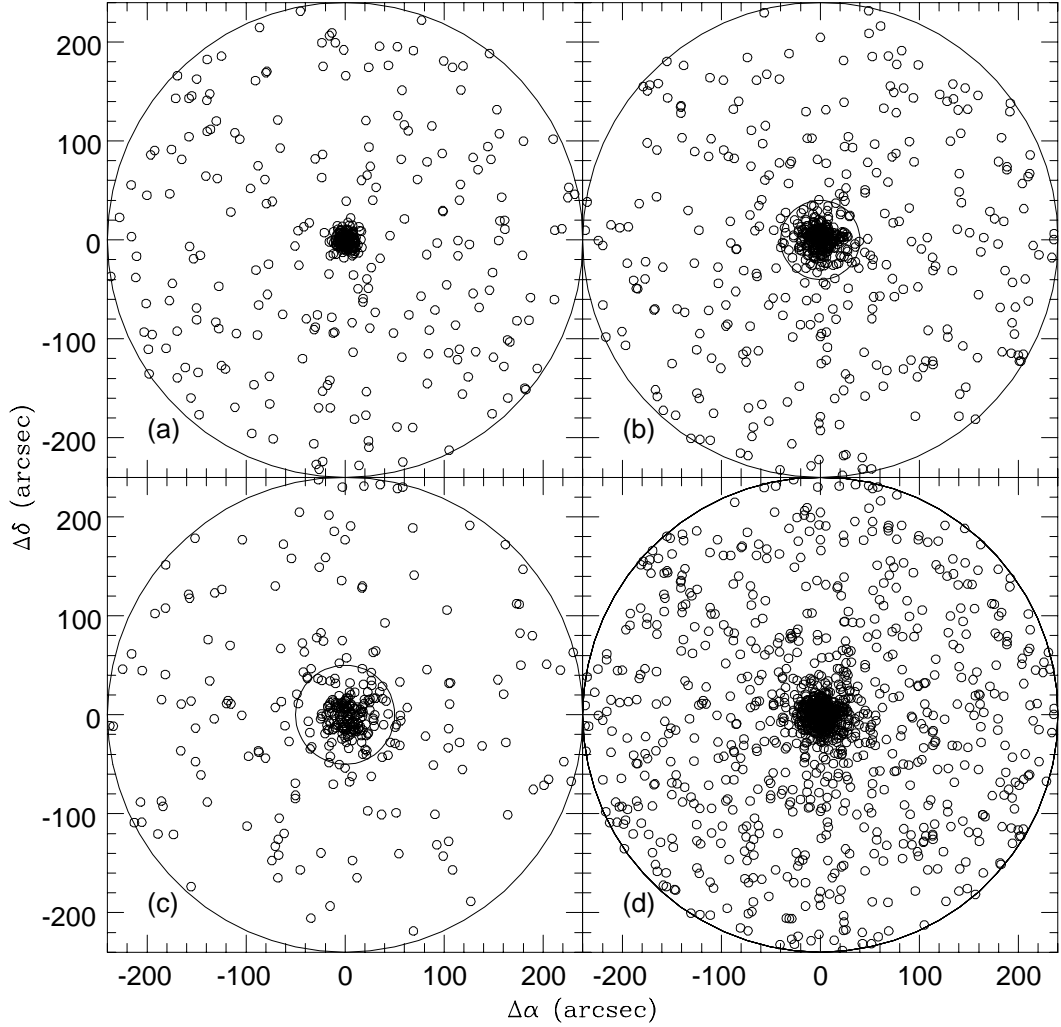}
\caption{The distribution of the offsets between 
the position of AGNs and stars in the V\&V96 and HIC catalogs
and the X-ray position from our catalog. The 
first three panels correspond to three different ranges 
of the offaxis angle in the ROSAT PSPC field: (a) from 0$\arcmin$ to 
19$\arcmin$; (b) from 19$\arcmin$ to 35$\arcmin$; (c) 
from 35$\arcmin$ to 47$\arcmin$. In (d) are reported all 
the 1485 correlations. The circles of radius $r_{90}$=14$\arcsec$, 
40$\arcsec$ and 50$\arcsec$ plotted in (a), (b) and (c) represent
the 90\% X-ray error circles derived from the analysis presented in the text. 
\label{err_x}}
\end{figure}

By using an impact 
parameter of 4$\arcmin$ we found 960 X/V\&V96 correlations and 
525 X/HIC correlations for a total of 1485 sources.
The distribution 
of the offsets in right ascension ($\Delta\alpha$) and 
in declination ($\Delta\delta$) between the X-ray and the V\&V96 or HIC 
positions is shown in Figure~\ref{err_x}. We have divided the sources in 
three groups, corresponding to different ranges of the offaxis angle 
($\theta$): (a) from 
0$\arcmin$ to 19$\arcmin$; (b) from 19$\arcmin$ to 35$\arcmin$; (c) 
from 35$\arcmin$ to 47$\arcmin$. 

Using two different methods, we have estimated from these data the positional 
errors of the X-ray sources as a function of the offaxis angle. 

In the first method we compute, from the three panels 
presented in Figure~\ref{err_x} the radius of the circle that includes 
90\% of the {\it real} positional correlations (the total correlations
minus the expected chance coincidences), i.e. 
the radius ($r_{90}$) of the circle which contains 90\% of the points is 
defined by:

\begin{equation}
\frac{N(\leq r_{90}) - N_{sp}(\leq r_{90})}{N_{tot} - N_{sp TOT}} = 0.90 
\end{equation}

where:

 $N(\leq r_{90})$ is the number of correlations with offset $\leq r_{90}$;

 $N_{sp}(\leq r_{90})$ is the number of expected spurious correlations with 
offset $\leq r_{90}$;

 $N_{tot}$ is the total number of correlations within the 
4$\arcmin$ radius;

 $N_{sp TOT}$ is the total number of expected spurious correlations within the 
4$\arcmin$ radius;

We have estimated $N_{sp}$ and $N_{sp TOT}$ by assuming that all 
the sources with an offset $>r_{sp}$ were of spurious origin.
The value of $r_{sp}$ ranges from 50$\arcsec$, for $\theta \leq$19$\arcmin$, 
to 100$\arcsec$, for 35$\arcmin\leq\theta\leq$47$\arcmin$. 
This analysis indicates that the error circles at the 90\%  confidence level
are: 14$\arcsec$ for $\theta<$19$\arcmin$, 40$\arcsec$ for 19$\arcmin$
$\leq \theta <$35$\arcmin$ and 50$\arcsec$ for 35$\arcmin\leq\theta\leq$ 
47$\arcmin$. 

Alternatively, we have studied the differential distribution of the 
number of correlations found, defined as: 

\begin{equation}
d(r) = \frac{N(r,r+\delta r)}{\pi\times[(r+\delta r)^2 - r^2]}
\end{equation}

where:

$N(r,r+\delta r)$ is the number of correlations with offset 
between $r$ and  $r + \delta r$;

$\delta r$ is the width of the bin used (=2$\arcsec$).  

Then, we have performed a fit to this distribution using a Gaussian profile,
assumed to represent the  real coincidences, 
plus a constant, representing the spurious matches. The 
radius of the 90\% error circle is then obtained by the $\sigma$ of the 
Gaussian fit. The problem related to this method is that the observed
profiles typically are not  Gaussian, showing a systematic excess (wings) 
at large values of $r$. Moreover, the first bins contain few objects 
so that the uncertainties are large. By excluding the first 3 bins
from the analysis we have found values of the $r_{90}$ in good agreement
with those estimated with the first method. 

In conclusion, for the positional cross-correlation we use an X-ray  
error circle of radius 14$\arcsec$, 40$\arcsec$ and 50$\arcsec$ for 
$\theta<$19$\arcmin$, 19$\arcmin\leq \theta <$35$\arcmin$ and 
35\arcmin$\leq\theta\leq$ 
47$\arcmin$ respectively.

\section{The REX selection}
\subsection{Criteria for the cross-correlation}
Given the estimate of the X-ray and radio positional error circles, 
the impact parameter ($b$) to be used during the cross-correlation 
to achieve the 90\%  completeness level
can be determined simply by summing quadratically the 
two positional uncertainties 
($\sigma_R$ and $\sigma_X$) corresponding to the 90\% confidence level
for the radio and the X-ray positions, 
i.e.:
\begin{equation}
b = \sqrt{(\sigma^2_R + \sigma^2_X)} \label{b}
\end{equation}
In practice, $\sigma^2_R \ll \sigma^2_X$ (see Condon et al. 1998 
and \S~ 4.3) and thus $b \sim \sigma_X$.
Since we use the 90\% confidence level error circles,
we expect to lose about 10\% of the REXs using the impact parameter given by 
equation~(\ref{b}). This represents a compromise between reliability and 
completeness since a larger value of $b$ would increase the number of spurious 
radio/X-ray correlations while a smaller value of $b$ would increase 
the number of missed REXs. Clearly, the loss of 10\% of the REXs has to be 
considered when deriving the statistical properties 
like the LogN-LogS or the luminosity function. 
 
\begin{figure}
\plotone{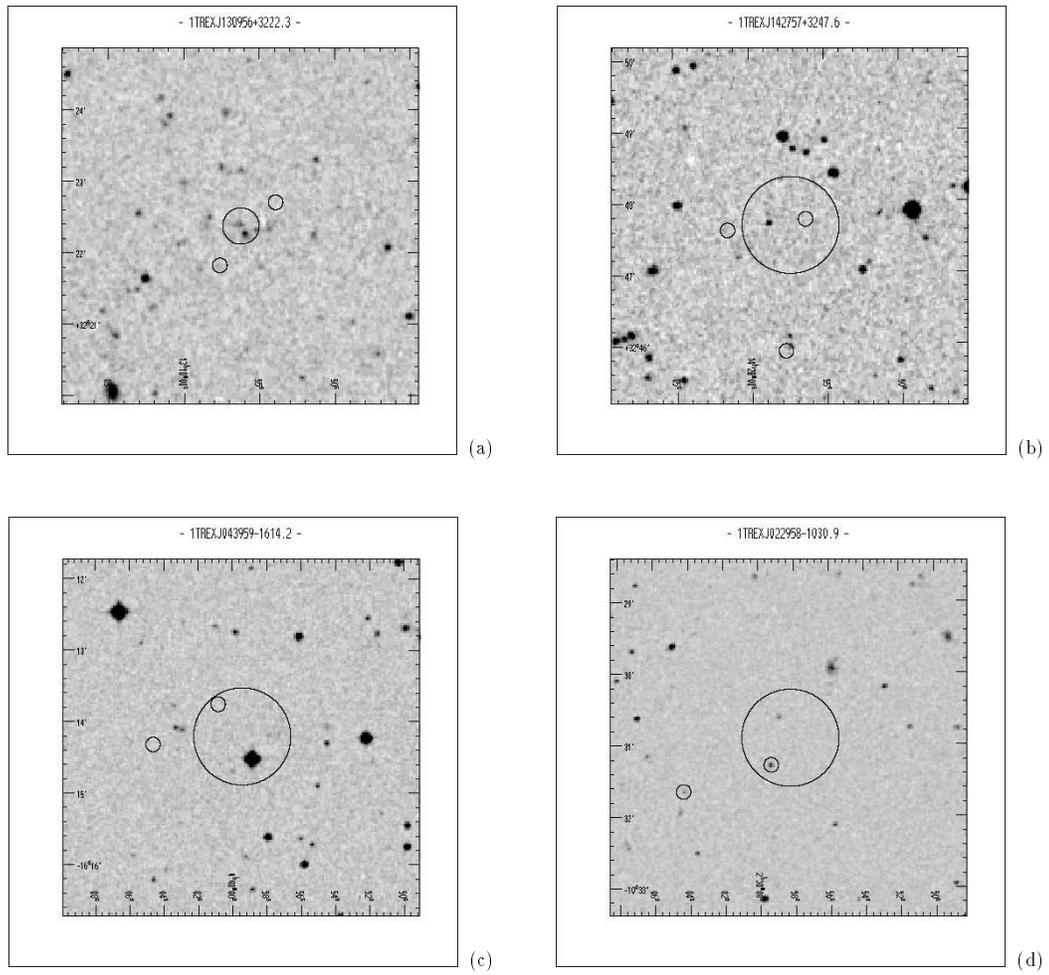}
\caption{(a) An example of REX that would be totally lost in a ``blind''
cross-correlation between radio (small circles) and X-ray (large circle) 
positions; 
(b) An example of an extended radio source which may introduce
an erroneous identification of the optical counterpart; 
(c) An example of an extended radio source which may induce  
a spurious radio/X-ray correlation; 
(d) An example of ``false'' double radio source. 
\label{extend}}
\end{figure}

\begin{figure}
\plotone{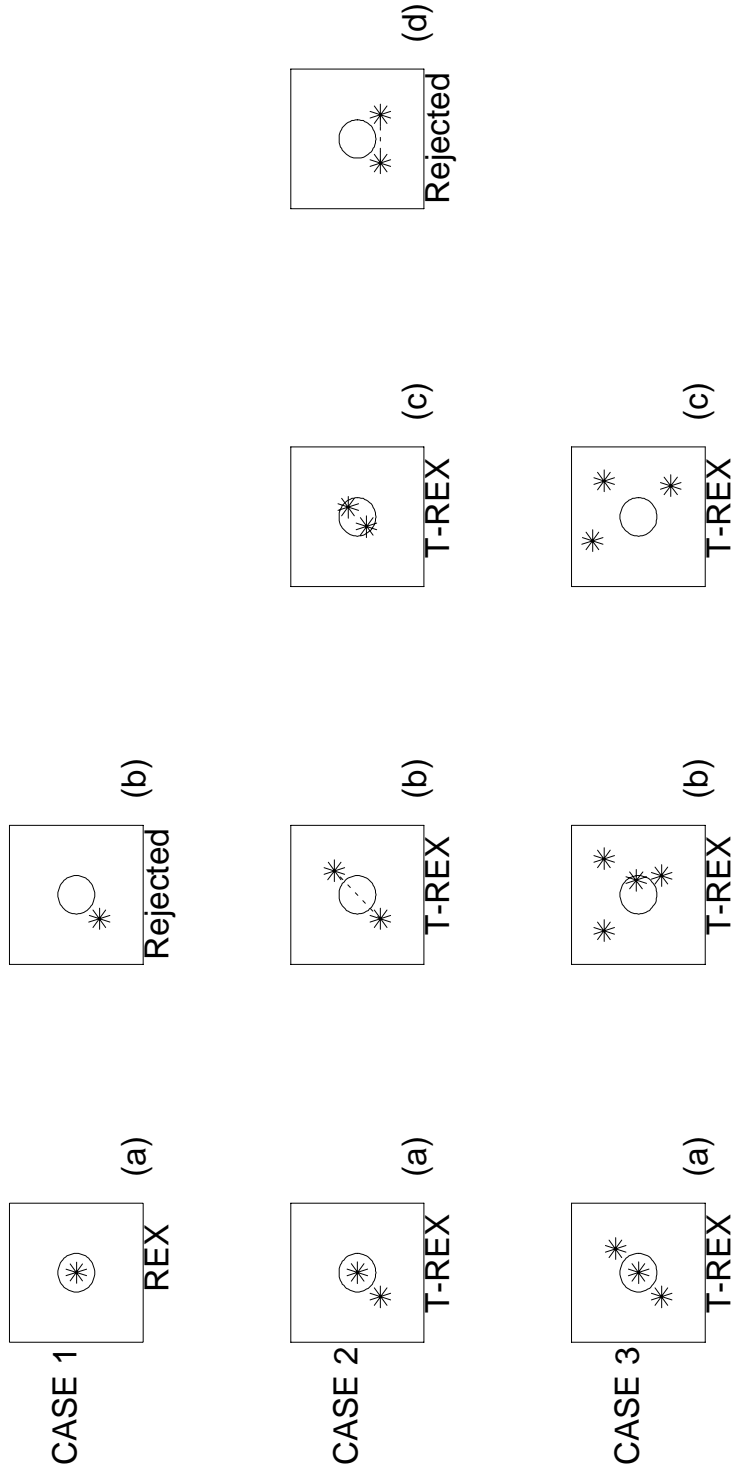}
\caption{Possible configurations resulting from the cross-correlation 
between radio sources
(indicated with stars) and X-ray sources (small circles), using an impact 
parameter of 2.5$\arcmin$ (box). There are three possible 
cases: 1 - only one radio source falls closer than 2.5$\arcmin$ 
to an X-ray source; 2 - two radio sources fall closer than 2.5$\arcmin$ 
to a  X-ray source; 3 - three or more sources fall closer than 2.5$\arcmin$ 
to an X-ray source. \label{cases}}
\end{figure}

A positional cross-correlation between the radio and the X-ray catalogs 
using the impact parameter $b$ given in equation~(\ref{b})  is the correct 
way to 
find REXs which are unresolved in the radio band. 
However, a large number of radio sources (e.g.  
radio galaxies and radio loud QSO) show characteristic extended 
structures (radio lobes, cores, jets) often well resolved in VLA images, 
even with the compact D and DnC configuration used for the NVSS.  
In these cases, the X-ray emission is centered in correspondence of 
the radio core and may be distant from the emission of the other 
structures (radio lobes). In 
a number of cases only the emission produced by the lobes is visible 
on the radio map and the object appears as a double radio source. As a 
consequence, several REXs (typically quasars and radio galaxies) could be 
lost in a ``blind" positional cross-correlation. 
The absence of a radio core in radio 
loud QSO is a well-know possible cause of incompleteness which may affect 
in particular the highly resolved VLA surveys. 
For example this effect can be one of the principal sources of incompleteness 
 for the FIRST bright QSO survey (Gregg et al. 1996). The existence of 
double radio sources may introduce also an erroneous identification of the 
optical counterpart of the REXs in the case in which only one radio 
component falls inside the X-ray error circle: in this case the REX 
is not lost but the optical counterpart, if it is supposed to be coincident 
with the  radio position, could be misidentified or could be spurious. 
In Figure~\ref{extend},  four cases of two NVSS components 
(small circles) close to the X-ray position (large circle)
are presented. 
We note that the size of the X-ray error circles depends on the offaxis 
angle, as described in section~4.3, while, for clarity, the radio error 
circles are all plotted with a fixed size of 6$\arcsec$. 
Figure~\ref{extend}a is an 
example of a possible double radio galaxy whose lobes lie 
just outside a  X-ray error circle, which contains the optical galaxy.
If we cross-correlate using the impact parameter ``b'' defined above 
(eq.[\ref{b}]) this source would be lost. Figures~\ref{extend}b and c 
are similar cases of possible double radio galaxies in which one of 
the two components lies inside the X-ray error circle. In both cases the 
source would be selected as a REX but a search for the optical counterpart 
at the radio position of the component consistent with impact parameter 
``b'' would fail to produce the correct identification. In the case of 
Figure~\ref{extend}b the REX is correctly retained in the sample, 
while in Figure~\ref{extend}c the match would be spurious since the optical 
counterpart of the radio source is not consistent with the X-ray 
position. Finally, Figure~\ref{extend}d shows a false case of double 
radio source: one of the two components is a radio source 
consistent with X-ray position, while the second one is an unrelated radio
source. Given this complex situation, instead of cross-correlating 
with the parameter ``b'' we use the following procedure.  
We cross-correlate the two catalogs using a large impact 
parameter (2.5$\arcmin$); this yields a large list of positional 
coincidences, most of which of spurious origin. 
At this stage, double radio sources are not missed (if their sizes 
are less than $\sim 5\arcmin$), but we have 
to deal with the problem of separating the real coincidences from the 
spurious ones. 
The  possible cases can be divided in three categories schematically 
represented in Figure ~\ref{cases}: 

Case 1:  only one radio source falls closer than 2.5$\arcmin$ to an X-ray 
source;

Case 2: two radio sources fall closer than 2.5$\arcmin$ to an X-ray source;

Case 3: three or more radio sources fall closer than 2.5$\arcmin$ to an 
X-ray source.

The first case is the simplest to deal with: if the radio and X-ray positions 
are consistent within the impact parameter defined by equation~(\ref{b}) we 
consider this object a REX; otherwise we eliminate the source.

We call the other two cases 
T-REX, for {\it Temporary REX}. The number of T-REX is very large, because of 
the large impact parameter used for the cross-correlation, and, for 
this reason, we filter automatically some obvious cases 
of chance coincidences. In particular, we can 
exclude ``a priori'' situations in which 
the line joining the two radio positions does not intersect 
the X-ray error circle, i.e., the optical counterpart of the presumed double
source falls outside the X-ray error circle
\footnote{
In principle this criterion could exclude from the 
sample a fraction of the sources where the optical 
counterpart (or the radio core) is not aligned with the two radio lobes 
(e.g.  Wide Angle Tail - WAT - sources). 
However, the number of X-ray emitting WAT sources is not large:
by performing the cross-correlation 
using a larger  
tolerance (e.g. 1.5$\times$b) we have not found any evidence of 
possible Wide Angle Tail source. Therefore 
we are confident that this bias is not important in the REX sample.
Moreover we remind that complex radio morphologies, i.e. sources with 
three or more NVSS components, are all retained as T-REX and visually 
inspected afterward.}
(see Figure ~\ref{cases}, case 
2d) or, if it intersects the X-ray error circle, both the radio components 
are on the same side.
By excluding all these situations (in case 2) we reduce significantly 
the number of 
T-REXs which have to be considered. 
For the remaining cases,  we inspect 
the 5$\arcmin \times$5$\arcmin$ optical finding chart from the 
Digitized Sky Survey (DSS)\footnote{The Digitized Sky Survey is 
produced at the Space Telescope Science 
Institute (STScI) under U.S. Government grant NAG W-2166}
and the 5$\arcmin \times$5$\arcmin$ radio map produced from 
the NVSS images and we try to detect all the possible cases of 
radio extended REXs and to recognize 
sources that are clearly chance coincidences (considering the 
position of the plausible optical counterpart and the morphology 
of the radio emission). 
We have analyzed the sample of $\sim$ 1000 T-REXs resulting
from the cross-correlation and we
have found that, on the basis of the radio and optical maps, we are able 
to distinguish a real REX from a chance coincidence in 
$\sim$80\% of the cases. 
The other 20\% T-REXs will be correctly classified only after 
gathering further information including the spectroscopic data for the 
possible optical counterparts. 
Of the T-REXs that we have already classified 90 are real double or 
triple 
radio sources (see Figure~\ref{extend}a), 443 are represented by ``normal'' 
REXs, i.e. 
the radio source consistent with the X-ray position is not related 
to the other radio component(s) found within 2.5$\arcmin$ (see 
Figure~\ref{extend}d) 
and  318 are chance coincidences i.e. the radio and the X-ray sources are
not related. 

The source catalog containing the relevant information on the REXs (position,
X-ray and radio flux and relative errors, etc.) will be made available 
electronically
\footnote{
A status report on the REX project (README.REX)
can be retrieved by ``anonymous" ftp at {\it ftp.brera.mi.astro.it}
in the directory ``/pub/REX".}
as soon as the NVSS data become stable and the REX quality check is completed. 
At the time of this writing the NVSS catalog is still an evolving entity 
because
of the existence of a few snapshot fields lacking total intensity images
(corresponding to about 5-7\% of the area covered by the X-ray data). These
gaps are being filled by the NVSS group either by re-processing of existing 
data or by new observations.
We anticipate that we will release a REX catalog obtained using the latest 
NVSS version
and thus the total number of sources will differ marginally from that 
presented in this paper.
Furthermore, for each REX we intend to make available an optical finding chart
derived from the DSS.

\subsection{Estimate of the number of radio/X-ray  chance coincidences}

We can evaluate the number of chance coincidences  
produced by  the X-ray/radio cross-correlation in two different ways. 
One possibility is to calculate the number of spurious REXs 
analytically, i.e. computing  
the number of radio sources expected by chance within a circle of radius 
``b'' centered on an X-ray position. 
If $N_1$, $N_2$ and $N_3$ are the number 
of X-ray sources with $\theta<$19$\arcmin$, 19$\arcmin\leq\theta<$35$\arcmin$ and 
35\arcmin$\leq \theta \leq$ 47$\arcmin$, respectively, 
and $b_1$, $b_2$ and $b_3$ are the corresponding impact parameters used 
for the cross-correlation, the total number of expected
chance coincidences is :
\begin{equation}
N_{chance} = (N_1 \times \pi b_1^2  + 
N_2 \times \pi b_2^2  + 
N_3 \times \pi b_3^2)\times \rho
\end{equation}
where $\rho$ is the number  of radio sources per unit  area 
with a radio flux greater than 5 mJy. 
We have $N_1$=4943, $N_2$=7674, $N_3$=3658 and $\rho \sim$31 deg$^{-2}$ 
and, consequently, $N_{chance}$ = 170, that is about 10\% of the total 
number of the REXs. 

Alternatively, the number of chance coincidences can be 
derived by cross-correlating the two catalogs after applying a
 relative shift in position to one catalog: if this off-set is 
significantly greater 
than the impact parameter b, we expect 
that all the results of the cross-correlation are pure chance coincidences.
By using this second approach we have derived that the expected percentage
of chance coincidences in the REX sample is 11.5$\pm 2$\%, 
in agreement with the result of the other (independent) method. 

For what concerns the double or triple REXs, the computation of the spurious
matches is not so straightforward as for the ``single component'' REX.
On the other hand, these complex sources represent a small fraction 
(about $\sim$6\%) of the total number of REX and thus the 
overall percentage of chance coincidences is practically determined by the 
``single component'' REX. 

Finally, we note that the estimated percentage of chance coincidences 
is a function of the offaxis angle, since the size of the impact
parameter used for the cross-correlation depends on the position of
the source in the PSPC field. If we consider the three ranges of offaxis
angle independently, we find that the percentage of chance coincidences
is 2.6\% in the inner part of the PSPC field ($\theta\leq$19$\arcmin$),
12\% between 19$\arcmin$ and 35$\arcmin$ and 16\% for $\theta\geq$35$\arcmin$. 
Thus, upon completion of the identification program, a comparison of the 
identification content as a function of the off-axis angle will provide 
information useful for the recognition of the spurious matches. 

Moreover, if, during the identification process, we find within the X-ray
error circle a source which is a good candidate as optical counterpart
of the X-ray source but which is not positionally consistent with 
the radio source, we flag this REX as a possible spurious match. 

\subsection{Expected composition of the REX sample}

The expected composition of the REX sample depends on the radio, X-ray and
optical limits. We estimate that the main constituents
of the sample will be RL AGNs and BL Lac objects. Nevertheless,
a number of radio quiet (RQ) AGNs is 
also expected if their magnitudes are sufficiently bright. For instance, 
a RQ AGN with $\alpha_{ro}$ = 0.3 will be included  
in the REX survey (i.e. it will have a radio flux $\geq$ 5 mJy) 
only if its magnitude is brighter than $\sim$ 18. 
Also clusters of galaxies will be present in the REX sample, if they
contain at least one bright ($\geq$ 5 mJy) radio galaxy. In this case the 
term ``REX"
is somewhat improper because the radio and the X-ray emissions come
from physically disjointed (although related) sources 
(the hot diffuse plasma) in the X-ray band, and the radio galaxy, in the
radio band. The number of cluster of galaxies expected 
in the REX sample is difficult to compute. First of all, there is 
a strong bias against clusters which do not contain bright radio galaxies. 
Moreover, even in the case of
a cluster containing a powerful radio galaxy,
the position of such radio source has to be consistent, within the 
impact parameter ``b'', with the position of the center of the X-ray source, 
that has no relation with the physical dimension of the cluster.  
Since neither the fraction of clusters
containing a bright radio galaxy  nor 
the positional distribution of such radio galaxies within
the cluster itself are known accurately, a quantification of 
these biases is hard to compute. 
We stress that these
considerations imply  that the sample of clusters of galaxies
selected in this project can not be used for
statistical purposes. On the other hand, the REX sample could be very
useful to find individual high redshift clusters of galaxies.
For the same reasons discussed above, it is difficult to predict the
expected number of radio galaxies in the REX survey  since a significant
fraction of them is found in clusters. 
Thus, in this section, we will assess the 
predicted properties only of the AGNs and BL Lacs.  
To this end, we have used the current knowledge of the
statistical properties of these classes of objects (i.e. 
luminosity functions, evolution, X-ray - optical - radio relationships) .

{\it RL AGNs}. 

\begin{figure}
\plotone{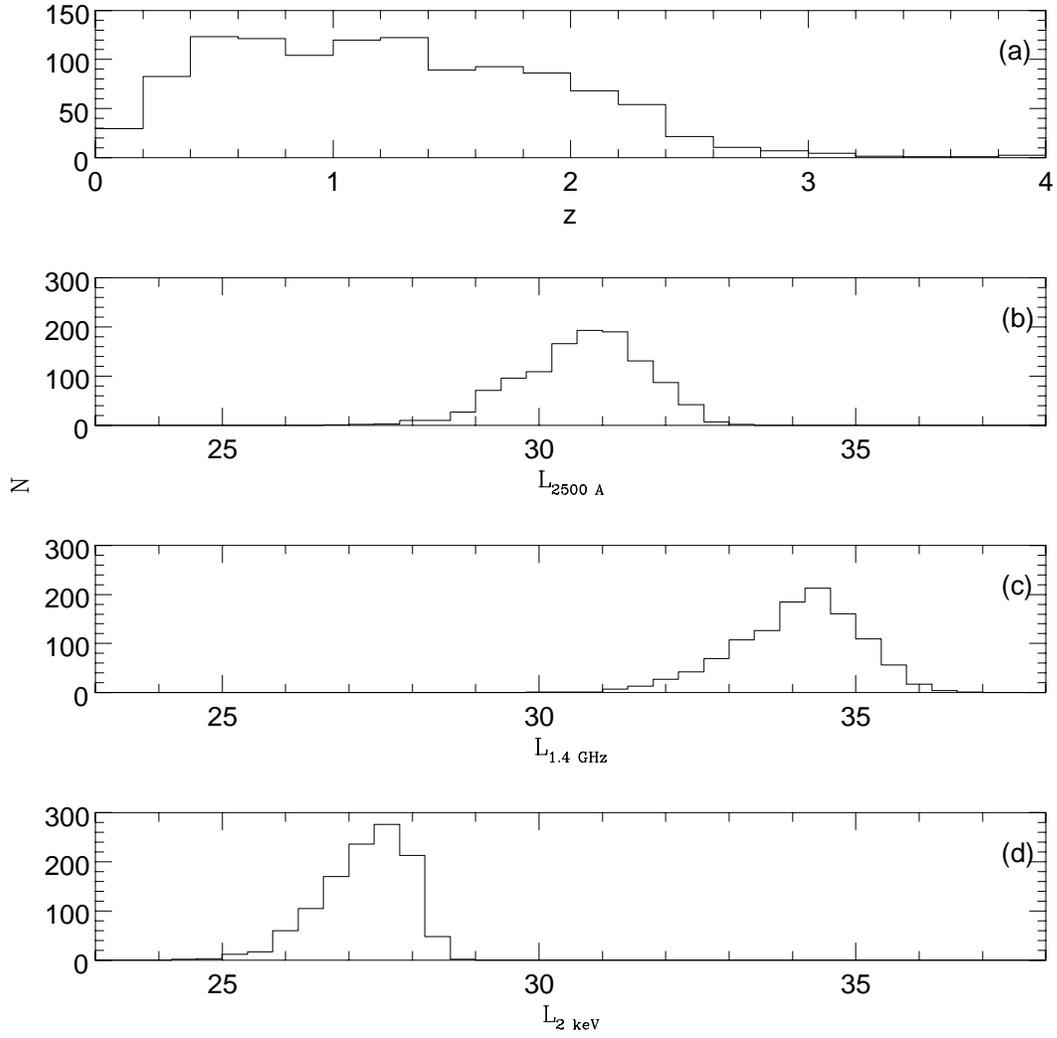}
\caption{The distribution of redshift (a), 
optical (b), radio (c) and X-ray (d) monochromatic luminosities 
of the simulated sample of RL AGNs. \label{qsr_sim}}
\end{figure}

The X-ray luminosity function (XLF) and the cosmological evolution of 
this class of objects have been adapted from Della Ceca et al. (1994).  
The XLF
has been first rescaled from the 0.3-3.5 keV energy band to the 0.5-2.0
keV energy band and then integrated over the luminosity range $10^{42}$ to
 $10^{47}$ erg s$^{-1}$ and from z=0 to z=4.  We
have assumed the best-fit evolutionary model until z=2 and no evolution
at higher redshift. Given the shape of the XLF (flat at low luminosities 
and steep at high luminosities) the results of the integration are not 
particularly sensitive to the choice of the luminosity limits. 
By convolving the results  with the X-ray sky-coverage 
 we have derived a synthetic sample of sources characterized by 
an X-ray flux and a redshift.
Then, we have assumed a relationship between radio and X-ray
luminosities and between optical and X-ray luminosities, and we have
associated to each object the corresponding radio and optical flux.
To this purpose, we have used the correlation between the total
(extended plus core) radio luminosities at 1.4 GHz and the luminosities
at 2 keV and the relationship between the luminosities at 2500 \AA \ 
and the luminosities at 2 keV, derived by Brinkmann et al. (1997) for
the RL QSO detected in the RASS. 
We have added a ``synthetic'' noise to the relationships to mimic 
the spread of the observed distributions. 
Finally, we have imposed  the radio limit (flux$_{1.4 GHz} \geq$ 5 mJy), 
obtaining an expected sample of $\sim 1200$ RL AGNs. As previously discussed 
(section \S~5.1), we expect to loose about 10\% of real REX because of the 
impact parameter
used in the cross-correlation. As a consequence, the predicted number of 
RL AGNs in the REX sample is about 1100. 
In Figure~\ref{qsr_sim} we show the distributions of radio, X-ray and
optical  monochromatic luminosities and the redshift distribution for this 
sample. 

{\it BL Lac objects}

\begin{figure}
\plotone{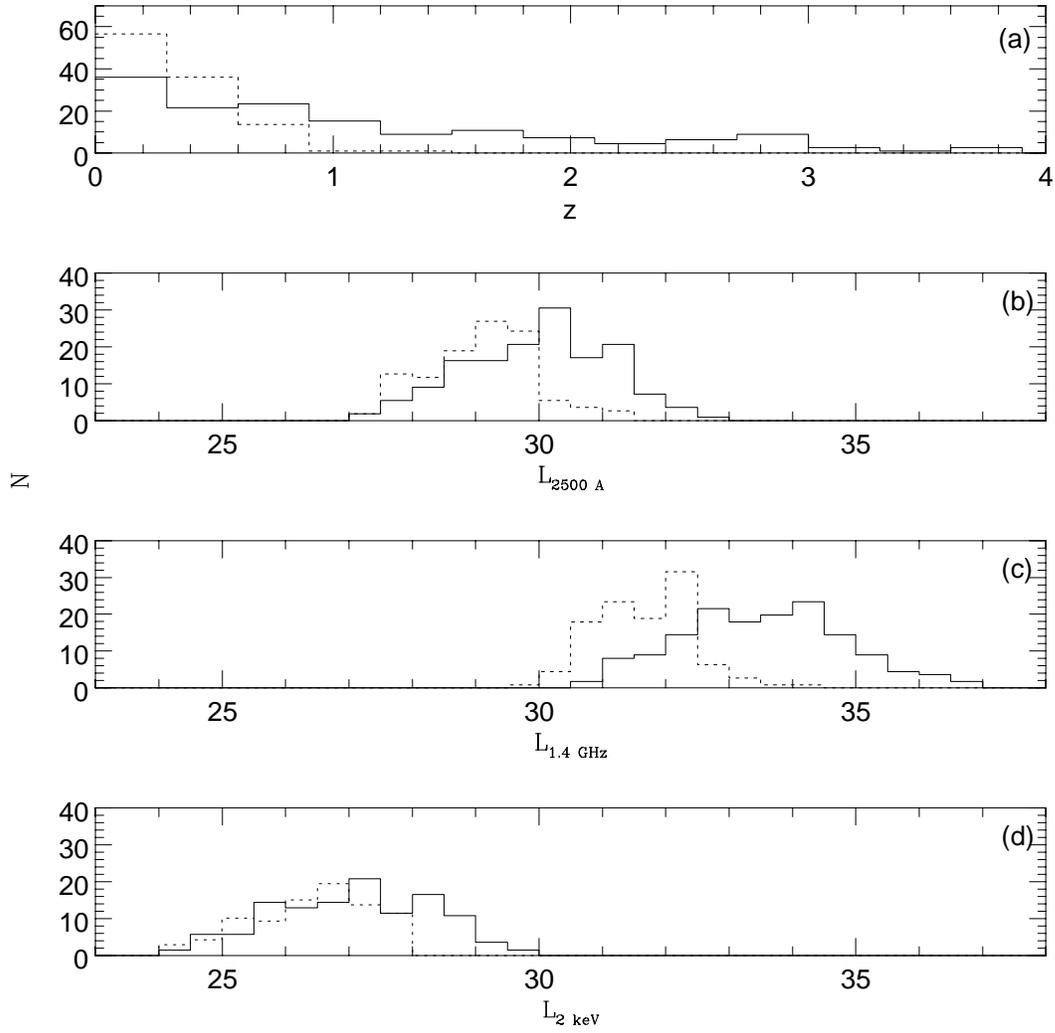}
\caption{The distribution of redshift (a), 
optical (b), radio (c) and X-ray (d) monochromatic luminosities 
of the simulated sample of BL Lacs, divided into ``RBL-type'' (continuous 
line) and ``XBL-type'' (dashed lines). \label{bl_sim}}
\end{figure}

We have considered the two population of BL Lacs (i.e. XBL-type and
RBL-type, see section \S~8.2) separately.  For both the XBLs and the RBLs we have used the
X-ray LFs and the evolutionary parameters presented in Wolter et al.
(1994).
After rescaling the XLFs to the 0.5-2.0 keV energy band, we have
integrated them from 10$^{42}$ to 10$^{47}$ erg s$^{-1}$  and from z=0
to z=4. We have assumed no-evolution after z=2.
From Wolter et al. (1994) we have also used the relationships between
the radio and X-ray luminosities and between the optical and X-ray
luminosities. Then, as for the RL AGNs, we have imposed the radio limit
(flux$_{1.4 GHz} \geq$ 5 mJy) to derive the expected sample of BL Lacs. 
We have obtained a simulated sample
of about 290 BL Lacs, 120 of which are XBL and  170 are RBL. Taking into
account the 10\% of BL Lacs that will be lost during the cross-correlation,  
the effective number of BL Lacs expected in the REX sample is about 260. 
The properties of the simulated sample of XBL-type and RBL-type objects
are presented in Figure~\ref{bl_sim}.
We note that these results are sensitive to the evolutionary parameters and
to the luminosity ranges used in the integration of the luminosity
functions. Of course, the dependence of the results on the  
parameters used implies that the final composition of the REX sample will
provide strong constraints on the evolutionary properties of BL Lacs 
as well as on XBL/RBL ratio. 

{\it RQ AGNs.}

As stated previously, we do not expect a high number of RQ AGNs
($\alpha_{ro} <0.35$) in the REX sample because of the presence of a
relatively high radio flux limit.
For a comparison, in the EMSS sample (Maccacaro et al. 1994) only 6 of the
382 RQ AGNs ($\sim$2\%) have a radio flux (at 5 GHz) above 5 mJy and 
most are not detected even at 1 mJy.
To quantify the number of expected RQ AGNs, we have used
the XLF and evolution
model of the RQ AGN population as given in Della Ceca et al.
(1994), adapted to the 0.5-2.0 keV energy band, and we have integrated the 
derived XLF  from $L_X = 10^{41}$ to $10^{46}$ erg s$^{-1}$ and from
z=0 to z=4.  We have assumed the best-fit evolutionary model until
z=2 and no evolution afterwards.
The radio (or radio upper limit) and optical flux to be associated to
each simulated X-ray source has been computed under the hypothesis
that the $\alpha_{rx}$ and $\alpha_{ro}$ distributions of the
RQ AGNs in the EMSS sample are representative of the class.
The expected number of RQ AGNs in the REX survey obtained after imposing the
radio constraint is $\lesssim$ 100.

\begin{figure}
\plotone{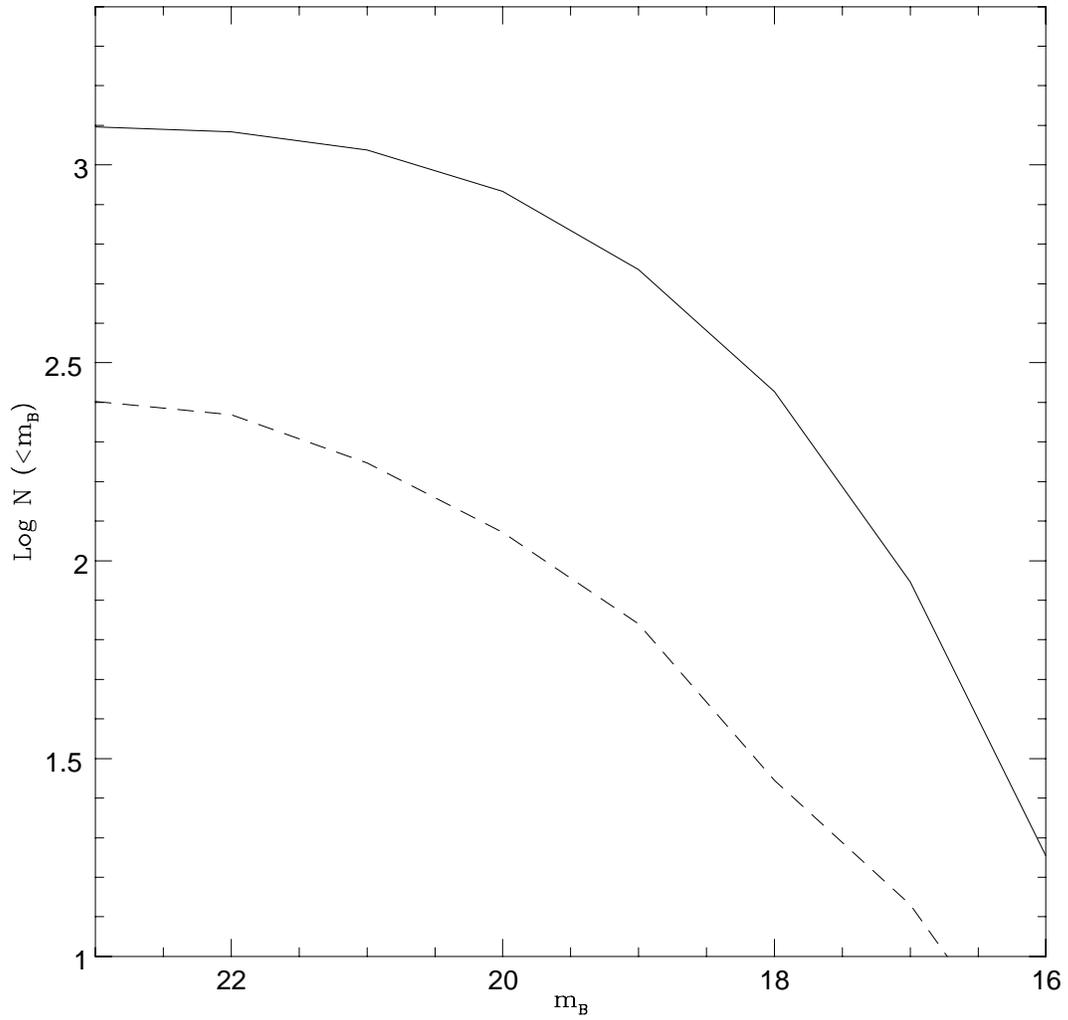} 
\caption{The logarithm of the expected number of AGNs, both radio loud and 
radio quiet (continuous line) and 
BL Lac objects (dashed line) present in the REX sample, as a function of 
the B magnitude. \label{compos}}
\end{figure}

In  Figure~\ref{compos} we show the total number of AGNs (both radio loud and 
radio quiet, solid line) and BL Lacs (``XBL'' + ``RBL'' type, dashed line) expected in the REX 
sample as a function of the limiting magnitude, $m_B$. We note that 
70 - 80\% of the expected BL Lacs and AGNs are brighter than $m_B = 21$, and 
thus observable with 4 meter-class telescopes. 

\section{Optical identification of REXs}
The positional accuracy of the VLA data guarantees that there is, in general, 
only one possible optical counterpart for each REX, at least for 
$m_B \leq 21$. 
In the case of a T-REX
the situation is slightly different but in general the objects to observe
spectroscopically are at most two. This represents a great advantage over 
similar
projects requiring optical identification of X-ray catalogs derived from PSPC
observations and makes the complete identification of the REX sample a feasible
endeavor. The importance of dealing with a fully identified sample of sources
is fundamental as it was clearly demonstrated by previous such surveys
(e.g. the {\it Einstein} EMSS).
Even with a VLA position, however, one should pay attention to the possibility
of finding spurious optical candidates falling within the error circle, in
 particular at
faint magnitudes. We have assessed the expected probability of finding 
interlopers
in  error circles of 5$\arcsec$ and of 2$\arcsec$ radius (95\% confidence for 
a NVSS source
of 5 mJy and $\geq$ 10 mJy, respectively) using the surface density of stars, 
galaxies and QSOs at the 
POSS II limit, as given by Condon et al. (1998). The results are extremely 
encouraging 
since the number of spurious objects of all kinds expected is of the order of 
few percent (radius of 5$\arcsec$) or $<$ 1\% (radius of 2$\arcsec$). 

For each REX  we create a 
$5\arcmin \times 5\arcmin$ finding chart from the 
DSS material and
we search for the probable optical counterpart. Using the 
NED facility we have identified a number of sources from the literature. 
At the same
time we have initiated spectroscopic observations at   
the 88$^{''}$ telescope of the University of Hawaii (UH) in Mauna Kea (USA),
at the 2.1m telescope 
of UNAM in S. Pedro Martir (Mexico) and at the 2.2m and the 3.6m  
telescopes of ESO (Chile). 

The spectroscopic observations have been carried out during the 
period 1995/1998 using a long-slit and low dispersion (from 3.7 \AA/pixel 
to 13.2 \AA/pixel) set-up to 
maximize the wavelength coverage. The details of the set-up used 
during the observing runs are summarized in table~\ref{runs}. Details 
about the individual sources  observed  
at the S. Pedro Martir 2.1m telescope  in the period 
April - September 1995 have been presented in Wolter, Ruscica \& Caccianiga 
(1998). 

For the reduction of the data we have used the IRAF {\it longslit} package. 
The spectra have been calibrated in wavelength by using an He--Ar (UNAM, ESO) 
or a Hg--Cd--Zn (UH)  reference spectrum. The photometric standard stars 
observed for the flux calibration are: Feige 34 (UNAM, UH), BD+284211 (UNAM), 
HD 19445 (UH), SAO 098781 (UH), LTT 377 (ESO) and HD 84937 (UH).
No attempt has been made at performing an absolute photometric calibration. 

In total, at the time of this writing, 
 we have observed and identified 125 REX. 
We have classified the objects on the basis 
of the features observed in their optical spectrum. 
We classify a source 
as {\it Emission Line object} if at least one strong (EW$\geq$5~\AA\  
in the source rest frame) 
emission line is present in the spectrum. Then, on the basis of the 
width of the observed line(s) we call the object  {\it Broad Emission 
Line AGN} if the FWHM of at least one emission line is greater than 
1000~km/s (in the rest frame of the source) or {\it Narrow Emission 
Line Object} if all the observed lines have FWHM $<$ 1000~km/s. In this last
case we have applied, when possible, the diagnostic criteria presented
in Veilleux \& Osterbrock (1987) to distinguish a starburst galaxy 
from an AGN. 

In the case of the objects without any strong emission line (EW$<$5~\AA) 
we have used the relative depression
of the continuum across $\lambda$=4000~\AA\ (the {\it Ca~II contrast}) 
as an indicator of the presence of a nuclear non-thermal component (a BL Lac 
nucleus) in the host galaxy. 
We have computed its amplitude following 
Dressler \& Shectman (1987), i.e. by estimating the average fluxes 
(expressed in unit of frequency) between 3750~\AA\  and 3950~\AA\ ($f^-$) and 
between 4050~\AA\  and 4250~\AA\  ($f^+$) in the rest-frame of the 
source; the contrast is then defined by:
\begin{equation}
K(Ca~II)  = \frac{f^+ - f^-}{f^+}
\end{equation}

\begin{figure}
\plotone{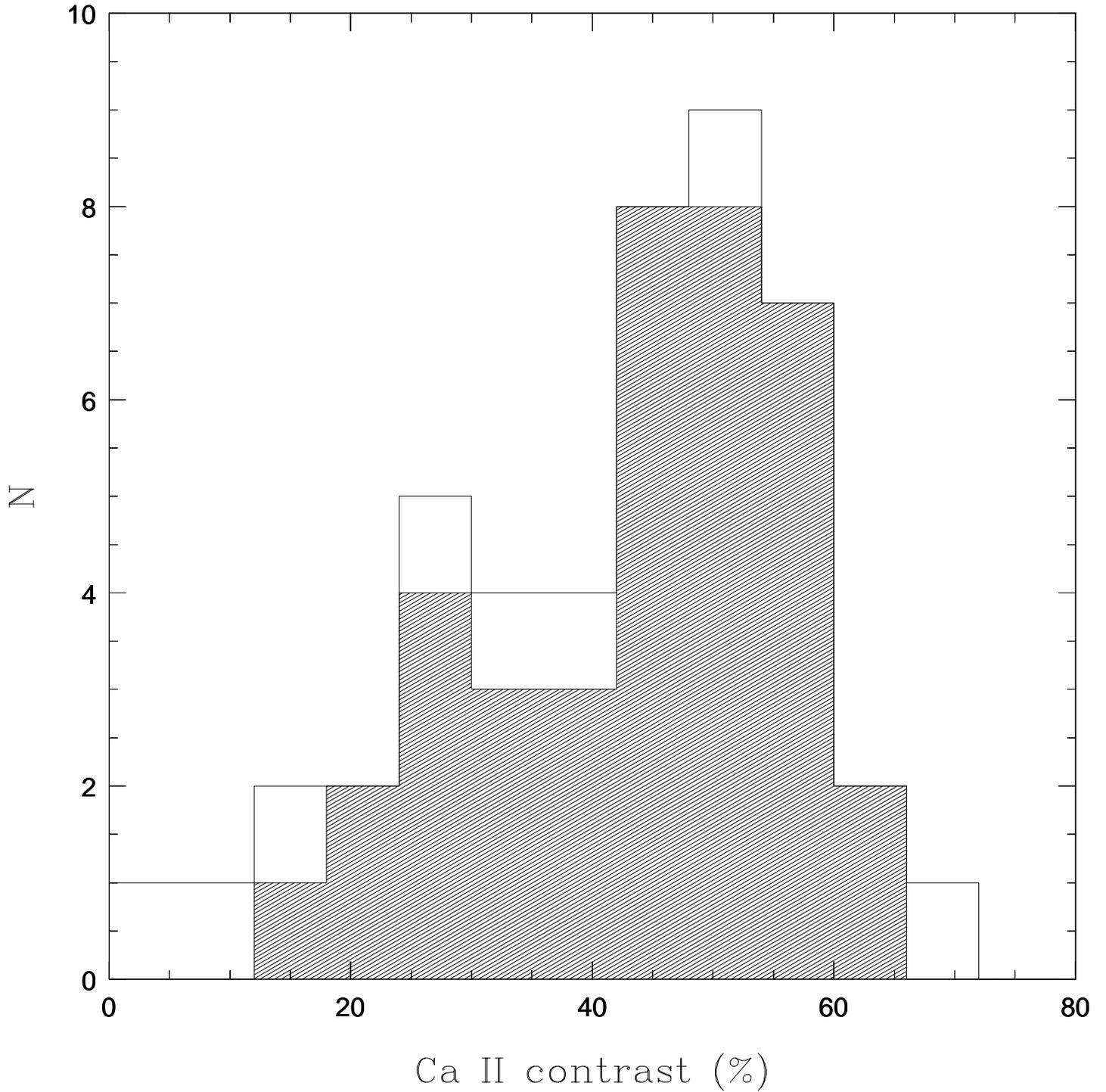} 
\caption{The distribution of Ca~II contrast  of the objects without 
optical emission lines discovered in the REX survey. 
The shaded histogram represents only the objects with a firm estimate
of redshift. 
\label{hist_ca}}
\end{figure}

If this feature is absent or if it has an 
intensity less than 25\%, the optical emission of the source is dominated 
by the non-thermal nucleus and we define the object as {\it firm BL Lac}
according to the usual definition (e.g. Stocke et al. 1991). 
This limit had been proposed on the basis of the study of the 
spectroscopic properties of a sample of (mostly) elliptical galaxies 
presented in Dressler \& Shectman (1987). Stocke et al. (1991) pointed 
out that only 1.2\% of the galaxies contained in the Dressler \& Shectman 
sample have K(Ca~II)$\leq$25\% (and no emission lines with  
EW$>$5~\AA) and that, consequently, it is extremely unlikely that normal 
elliptical galaxies would be classified by mistake as a BL Lac object, by 
using these definition criteria. Moreover, the adopted classification 
was supported by the absence, in the EMSS sample, of objects without emission 
lines (EW$<$5~\AA) and with a Ca~II contrast between 30\% and 40\% 
(see Figure~4 in Stocke et al. 1991). In the last years, however, after  
the discovery of a large number of new BL~Lacs from 
the identification of the ROSAT All Sky Survey catalog  (Laurent-Muehleisen et
 al. 1998) or from a sample of optically bright flat spectrum radio sources 
(March\~a et al. 1996), this ``gap'' has been partially filled up and the 
distinction between a ``normal'' elliptical galaxy and a BL Lac object 
has become blurred, at least in the optical band. Therefore 
it is necessary a re-definition of the classification 
criteria used to distinguish a BL Lac from an elliptical galaxy. 
To this end, we have studied the objects found in the REX sample without
optical emission lines and with a measurable Ca~II contrast. 
In Figure~\ref{hist_ca} we have reported the observed distribution
of the values of Ca~II contrast for this sample of sources. 
The histogram 
is peaked around K(Ca~II)$\sim$50\%\ ,that is consistent with the mean 
value  found by Dressler \& Shectman (1987) for a population of 
elliptical 
galaxies. However, unlike the Dressler \& Shectman sample, where only 5\% of 
galaxies have a K(Ca~II)$\leq$40\%, our sample
contains 18 objects out of 46 (39\%) with a K(Ca~II)$\leq$40\%.
This excess is statistically
significant ($\geq$5$\sigma$) and it is also present in the samples 
studied by Laurent-Muehleisen et al. (1998) and March\~a et al. (1996). 
We consider this excess as evidence of the presence, at least in the objects 
with K(Ca~II)$<$40\%, of an 
extra source of continuum that lowers significantly the values of the 
Ca~II contrast. Also the objects with K(Ca~II)$>$40\% could, in principle, 
harbour 
an optically weak BL Lac nucleus, whose intensity is not sufficient 
to reduce significantly the value of Ca~II contrast. We are not able to 
distinguish 
these cases using only our  optical discovery spectra. 
The lack of a sharp bimodality in the Ca~II contrast distribution makes
the problem of the definition of a BL Lac a task somewhat arbitrary (at least
using only optical spectroscopy). On the other hand, as discussed in 
section \S~8.1, the existence of objects intermediate between a typical 
BL Lac and a typical radio galaxy is of fundamental importance in the 
context of the ``beaming model" and should be investigated with attention. 

Nevertheless a limit on the value of Ca~II contrast to distinguish 
a BL Lac from a galaxy is needed, at least, from an operative 
point of view.
While it is now clear that the usual 
limit of 25\% fails to find a significant fraction of BL Lacs, a new
higher limit could classify as BL Lac some ``normal" elliptical galaxies. 
If we assume that all the objects in our sample
with K(Ca~II)$>$45\% are ``normal'' elliptical galaxies and 
that these objects are distributed, in terms of 
values of Ca~II contrast, as the elliptical galaxies studied by Dressler 
\& Shectman (1987) we expect that only 1 - 2 objects among those with 
a Ca~II contrast less than 40\% (18 in total) are ``normal" galaxies 
(i.e. $\sim$10\%). 
Therefore we call {\it BL Lac candidates}
all the objects without emission lines (EW$\leq$5~\AA) and 
25\%$<$K(Ca~II)$\leq$40\%.  
Further observations will be necessary to distinguish
the galaxies harbouring a BL Lac nucleus from ``normal'' galaxies. 
In  section \S~8.1  we will 
present a study of the X-ray luminosities of the BL Lac 
candidates that supports the hypothesis of the
presence of a non-thermal nucleus at least in a fraction of these objects. 

Finally, March\~a et al. (1996) proposed a revision also of the 
limit of 5~\AA\ usually imposed on the equivalent width of the 
emission lines present in the optical spectrum. Basically, they point out  
that the observed equivalent width of 
an emission line is relative to the total continuum that is 
composed by the thermal emission from the host galaxy plus the non-thermal 
contribution from the active nucleus. Since the non-thermal nucleus 
is strongly dependent on the 
viewing angle, they proposed to refer the intensity of the emission lines
to the galaxy starlight. This means that, given a defining limit 
on the equivalent width of the emission lines (relative to the starlight), 
the observed equivalent width (i.e. relative to the total emission) depends 
on the Ca~II contrast strength (see Figure~6 of March\~a et al. 1996).
We note that a criterium based on the
strength of any emission lines relative to the brightness of the host 
galaxy may introduce in the sample a bias since the same 
BL Lac nucleus could be differently classified on the basis of the 
luminosity of the elliptical galaxy that harbours it. 

In any case, at present, we have found only one object that fulfills the 
defining criteria proposed by March\~a et al. (1996)
(K(Ca~II)$\sim$30~\% and
an emission line - [OII]$\lambda$3727 - with EW$\sim$30~\AA). 
The spectrum of this source resembles that of a 
spiral galaxy Sb or Sc like NGC~4750 or NGC~6643 (see 
Kennicutt 1992). Thus, we define this object as a ``possible"  
BL Lac but we do not consider it in the analysis of BL Lac candidates 
presented here.

In summary, we classify an object  without strong optical emission
lines (EW$\leq$5\AA) as {\it ``firm" BL Lac}, if the Ca~II contrast is
absent or below 25\%; as {\it BL Lac candidate}, if the Ca~II contrast
is between 25\% and 40\%; as {\it elliptical galaxy} (usually a radio galaxy,
given the high luminosity in the radio band, as described below)
if the Ca~II contrast is above 40\%. 

We note that the definition of an object ``without emission lines" 
could be dependent on the observing set-up since the capability of  
detecting an emission line depends, 
among other thing, on the actual wavelength coverage. 
Our spectra cover the range between  $\sim$4000\AA\ and $\sim$8000\AA, 
extended, in some cases, blue-ward to 3500\AA\  and/or red-ward to 9000\AA. 
An object without emission lines in this range is, in general, 
a real featureless object. Nevertheless, some objects may show, 
in a low signal-to-noise spectrum, only a strong 
H$_\alpha$ (see e.g. MS0007.1$-$0231 in Stocke et al., 1991). 
If the redshift of the source is such 
that the H$_\alpha$ emission line falls outside the observed wavelength  range, 
the object could be erroneously classified as ``no emission line object". 
A fraction ($\sim$ 20\%) of the objects in the REX survey without 
emission lines have not been observed in the spectral region where the possible H$_\alpha$ is expected. 
The re-observation of these objects with an adequate wavelength coverage 
is planned. A similar follow-up will be 
necessary also for the objects classified from literature (NED).

\section{The composition and the general properties of the REX sample}

The spectroscopical observations carried out so far have lead to the identification of 125 REX.
Another 268 REX have been identified from the literature (NED). 
Among the 393 REX identified we have found: 202 Emission Line
AGNs (67 new), 136 galaxies\footnote{ 
With the term ``galaxy", we intend all the objects defined as such 
in literature as well as all the elliptical galaxies that follow the criteria 
presented in \S~6. Consequently, this is a quite heterogeneous class, 
including both ``normal" galaxies (spiral or elliptical) and 
radio galaxies without emission lines in the optical spectrum.
To obtain a more precise and homogeneous classification of these objects 
follow up observations are required.}
(32 new) and 55 BL Lacs or BL Lac candidates
(26 new). However, we stress here that 
the sub-sample of identified REX is not expected to reflect the 
global composition of the REX sample since it is not a well-defined,
representative subset. 

At present, the percentage of BL Lac objects in the REX survey is about 14\%, 
and it should be considered  only as indicative of the efficiency of  
the REX survey in finding BL Lacs. Altogether, this number is high if compared 
with the radio or the X-ray selections. By extrapolating this number, 
we predict a total of $\gtrsim$200 BL Lacs that is in fair agreement with 
the number predicted with our simulations (\S~5.3).

\begin{figure}
\plotone{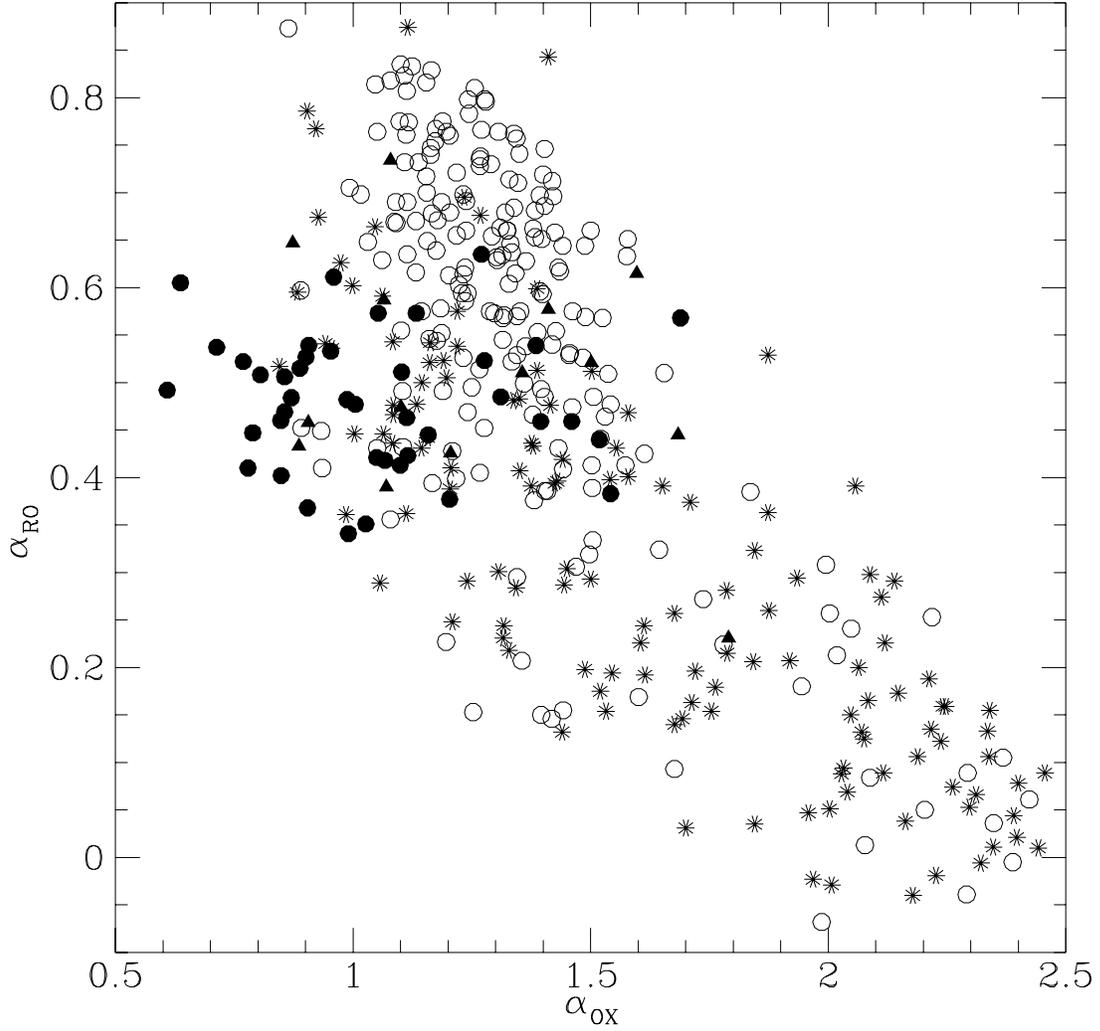} 
\caption{The radio-optical ($\alpha_{RO}$) 
versus the X-ray-optical ($\alpha_{OX}$) spectral indices of the 393 REX
identified. Emission line AGNs are represented as open circles, 
``firm'' BL Lacs as filled circles, BL Lac candidates as filled triangles and
galaxies as stars.  \label{arox}}
\end{figure}

In figure~\ref{arox} we have reported the radio-optical ($\alpha_{RO}$) 
versus the X-ray-optical ($\alpha_{OX}$) spectral indices of the 393 REX
identified so far. Emission line AGNs have been represented as open circles, 
``firm'' BL Lacs as filled circles, BL Lac candidates as filled triangles and
galaxies as stars. The values of $\alpha_{OX}$ and $\alpha_{RO}$
have been computed as in Stocke et al. (1991), using the monochromatic 
fluxes at 5~GHz, 2500\AA\ and 2~keV. 

As expected, the majority (84\%) of AGNs are radio loud ($\alpha_{RO}\geq$0.35)
but we find also a fraction (16\%) of radio quiet AGNs, selected 
thanks to the low radio flux limit. The percentage of radio quiet 
AGNs is high in comparison with our simulations (\S~5.3) but we recall that 
the sample of identified REX is not representative of the whole 
population and, in particular, it is strongly biased toward the bright 
magnitudes where the percentage of radio quiet AGNs is expected to be 
significantly higher. Galaxies 
are more widely distributed in the $\alpha_{RO}$/$\alpha_{OX}$ 
plane; on the other hand, in the case 
of galaxies in a cluster, a substantial fraction of the X-ray flux
probably comes from the intra-cluster gas. 

Figure~\ref{arox} shows that BL Lac objects occupy  a unique area 
in the $\alpha_{RO}$/$\alpha_{OX}$ plane, compared to the 
other classes of sources, as firstly noted by Stocke et 
al. (1991). Thus, a further selection in terms of $\alpha_{RO}$, 
$\alpha_{OX}$ could increase the efficiency of the search for BL Lac 
objects up to 25-30\%. On the other hand, a sample obtained 
in this way may be biased toward selecting the most ``extreme'' subset of the 
whole population.  

\section{The REX sample of BL Lacs: preliminary results}

Although the principal aim of this paper is to present the scientific 
goals and the selection criteria of the REX survey, two
preliminary results can be drawn from the analysis of   
the current sample of BL Lacs. Since this sample is not complete and 
representative, given the low rate of identifications, 
the conclusions must be considered as general indications of the 
potentiality of this survey. 

\subsection{The connection  between BL Lacs and Radio Galaxies}

In the context of the Beaming Model, BL Lac objects and FR I radio   
galaxies are thought to be the same class of sources seen at 
different viewing angles (e.g. Urry \& Padovani 1995). If
this picture is correct, we 
expect to find some transition objects with intermediate properties
between BL Lacs and FR~I radio galaxies. Up to now, this 
intermediate population of objects was missed, probably due to the 
limiting fluxes of the current X-ray/radio surveys and/or to the criteria 
adopted to classify a BL Lac. For example, 
in the Deep X-ray and Radio Blazar Survey (DXRBS, Perlman et al. 1998),
which is the result of a 
positional cross-correlation between the WGA catalog and a number
of radio catalogs, an additional constraint on the radio spectral index 
(i.e. $\alpha_R\leq$0.7, $f\propto\nu^{-\alpha}$) 
is imposed to further increase the efficiency of the selection. This 
constraint excludes from the DXRBS sample most of the ``normal'' radio 
galaxies. On the other hand, the presence of both radio galaxies and 
BL Lac objects in the same sample is of fundamental importance to 
address the problem of the existence of a population 
of ``transition'' objects between ``normal'' radio galaxies and BL Lacs. 
The existence of such population of sources is predicted by the 
beaming model and it is matter of discussion in recent literature 
(e.g. Browne \& March\~a 1993). 

In the REX survey we do 
not impose any further conditions other than the presence of the source 
in a radio and in an X-ray catalog. Moreover, as described 
above, we have developed a technique of cross-correlation that 
allows to select double radio sources (like radio galaxies) 
as well. For these reasons, we expect that the REX sample will contain 
both FR~I and BL Lacs as well as intermediate objects. 

In section~6 we have  
presented the discovery of low-luminosity
BL Lacs that show a value of Ca~II contrast intermediate
between BL Lacs and elliptical galaxies. To asses if these sources represent 
the 
connection between beamed objects (i.e. BL Lacs) and non-beamed objects 
(i.e. FR~I) we have studied the properties 
in the X-ray and in the radio band of all the objects for which 
we detect the Ca~II contrast in the optical spectrum (i.e. elliptical 
galaxies, BL Lac candidates and a fraction of BL Lacs). 

The information available for these objects comes from the 
ROSAT PSPC images and from the NVSS data used to select the REX sample. 

{\it Radio properties.}
The luminosities at 1.4~GHz of the BL Lac candidates and of the 
elliptical galaxies
newly discovered in the REX survey range from 10$^{31}$ to 10$^{33}$ erg 
s$^{-1}$ Hz$^{-1}$, consistent with those 
of FR~I and BL Lacs ($L_R \geq$10$^{30}$ erg s$^{-1}$ Hz$^{-1}$). 
Given this range of luminosity, the elliptical galaxies, that 
do not show any emission line in the optical spectrum, can be classified  
as radio galaxies, probably FR I.
For a fraction of objects we have information on the radio flux at 
5~GHz since they 
are included in the Green Bank catalog (Gregory et al. 1996) or in the 
PMN catalog (Wright et al. 1994). 
All the BL Lacs and BL Lac candidates present a slope 
$\alpha_R \leq$ 0.6 (f$\propto \nu ^{-\alpha}$) while the radio galaxies 
cover a wider range of values, from 
--0.2 to 1.2. The overall distribution
of the slopes does not reveal any bimodality. 

Radio galaxies and BL Lacs are expected to be different in terms 
of radio morphology. Typically, a radio galaxy with a radio luminosity 
between 10$^{31}$ and 10$^{33}$ erg s$^{-1}$ 
Hz$^{-1}$ and a redshift between 0.1 and 0.5 (like the objects discussed  
here) shows extended radio structures of 100 -- 500 kpc (Singal 1993) while 
BL Lacs show more compact radio morphologies (Perlman \& Stocke 1993; 
Kollgaard, Gabuzda \& Feigelson 1996). These differences cannot be 
well established with the NVSS data that are characterized by a large 
beam (FWHM=45$\arcsec$) that correspond to a linear size of 100 -- 300 kpc for 
redshift of 0.1 - 0.5. Specific observations with a better resolution 
are needed to study the differences between radio galaxies, 
BL Lacs and BL Lac 
candidates from the viewpoint of the radio morphology. For the sources
falling in the area of sky covered by the FIRST survey we will be able
to study the radio morphology with a resolution a factor 10 better than 
that achieved by the NVSS.

{\it X-ray properties}

The X-ray luminosities in the 0.5--2.0 keV band of the radio galaxies and 
BL Lac candidates are above 3$\times$10$^{42}$ erg s$^{-1}$. 
The objects with  K(Ca~II)$\geq$40\% (radio galaxies) 
show X-ray luminosities up to 10$^{44}$ erg s$^{-1}$ while some of 
the objects with K(Ca~II)$\leq$40\% (BL Lacs and
BL Lac candidates) reach luminosities of 5$\times$10$^{44}$ - 10$^{45}$ erg
s$^{-1}$
consistent (but at the low luminosity end) with the range of luminosities 
shared by BL Lac objects (10$^{44}<L_X<$10$^{47}$ erg s$^{-1}$). 
We recall that a typical FR~I 
galaxy (without emission lines in the optical spectrum) hardly reaches 
X-ray luminosities of 10$^{44}$ erg s$^{-1}$ (Fabbiano   et al. 1984). 
There are only few cases of objects (Silverman, Harris \& Junor 1998; 
Tananbaum et al. 1997; Caccianiga \& Maccacaro 1997) showing high X-ray 
luminosities 
($\geq3\times$10$^{43}$ erg s$^{-1}$) and an optical spectrum 
without any signature of the presence of an active nucleus (i.e. 
strong emission lines and/or a reduced value of the Ca~II contrast). 
There are strong evidences that these objects   
must be considered as BL Lacs whose optical emission
is overwhelmed by the light of the host galaxy. 

We note that some of the objects discussed here probably reside in clusters
of galaxies and, thus, the X-ray emission could be produced, in part or 
completely, by the intracluster gas. We have distinguished some of 
these cases on the basis of CCD observations of the optical field 
and/or on the basis of spectroscopic observations of 
other galaxies in the same field. 

\begin{figure}
\plotone{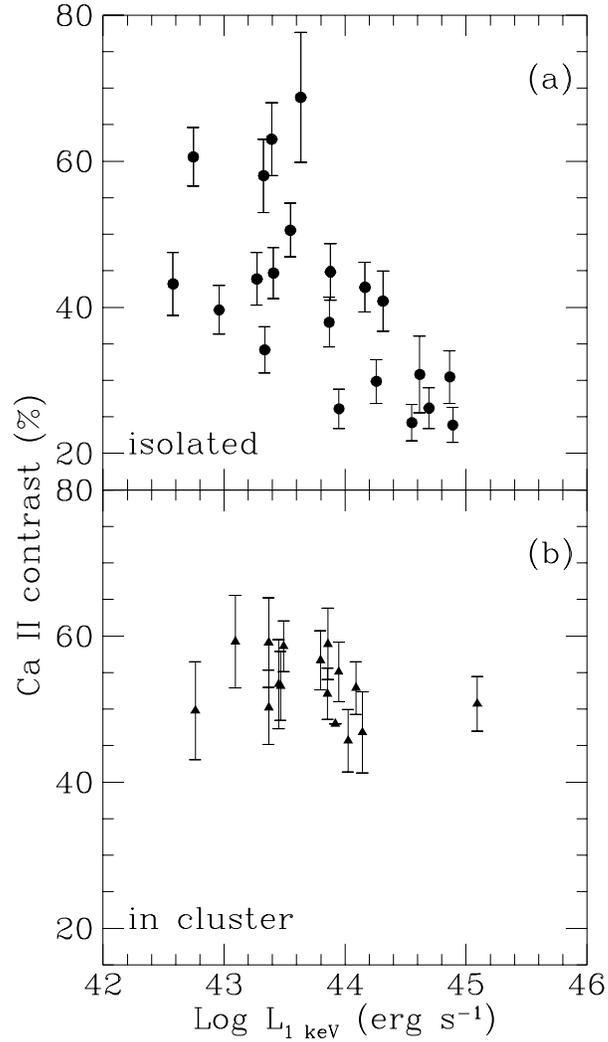} 
\caption{The value of Ca~II contrast versus the  monochromatic X-ray 
luminosity at 1~keV for the elliptical galaxies and BL Lac candidates 
newly discovered in the REX survey. 
We have distinguished, on the basis of the optical images, the objects for 
which there are no evidences for the 
presence of a cluster of galaxies (a) from possible clusters (b). Only 
the objects with a firm estimate of redshift have been considered.
\label{ca_lx}}
\end{figure}

In Figure~\ref{ca_lx} we have reported the values of Ca~II contrast 
versus the monochromatic luminosities at 1~keV for
the REXs for which we have no evidence of the presence 
of an overdensity of objects in the optical field (panel a) and 
for the REX in clusters (panel b).
Figure~\ref{ca_lx}a shows that the two quantities are anti-correlated 
(at $>$95\% confidence level), while clusters of galaxies 
(Figure~\ref{ca_lx}b) do not show any correlation. The correlation 
found for the isolated objects suggests a common origin of the high X-ray 
luminosity and the low value of the Ca~II contrast, probably 
related to the presence of a non-thermal nucleus in the host galaxy: 
the stronger is the non-thermal emission, the lower is the Ca~II 
contrast and the higher is the X-ray luminosity. In the case of clusters 
of galaxies, the strong X-ray luminosities are  mainly related to 
the intracluster gas and not to the activity of the nucleus and thus  
the values of Ca~II contrast and the X-ray luminosities are not expected
to be correlated. 

Figure~\ref{ca_lx} shows a continuity between the X-ray/optical properties 
of the radio galaxies and the BL Lac candidates discovered in the REX survey. 
This continuity suggests that the BL Lac candidates 
represents the connection between BL Lacs and their parent population.
In the context of the beaming model,
these sources could be less beamed than the ``usual'' BL Lac objects i.e. 
they are seen at viewing angles 
intermediate between FR~I ($\sim$90$^{\circ}$) and BL Lacs ($\sim$0$^{\circ}$ 
-- 30$^{\circ}$). 

We note that, given the modest angular resolution achieved by the PSPC 
instrument, in particular at large offaxis angles ($>$20$\arcmin$), 
we are not able to exclude the presence of a diffuse contribution to
the X-ray luminosity also for the objects for which we 
do not detect the presence of an overdensity of galaxies 
in  the optical field. 
More accurate and spatially resolved X-ray observations (e.g. with ROSAT HRI or 
AXAF) are needed in order to unambiguously asses the presence of a point-like 
source among the BL Lac candidates.

\subsection{XBL versus RBL}

Up to now, radio and X-ray surveys have selected different types of 
BL Lacs, called RBL and XBL, respectively. Typically, XBLs are less 
variable and less polarized than RBLs. The overall spectral distribution 
of the two classes is strongly bimodal and their cosmological 
evolution differs significantly (Morris et al. 1991; Stickel et al. 
1991; Wolter et al. 1994; Bade et al. 1998). Moreover, XBLs frequently show 
absorption features in their  optical spectrum similar to those of ``normal''
elliptical galaxies. This is an indication that the 
contribution of  the non-thermal emission, coming from the nucleus 
(or the jet), is not strong enough to cancel completely the stellar 
emission from the host galaxy. 
The existence of two distinct
populations of BL Lacs was tentatively attributed to a difference in the 
width of the emission cones of the radio and X-ray radiation (Ghisellini 
\& Maraschi 1989; Celotti et al. 1993): in this framework, 
the X-ray radiation is less beamed than the radio one and, consequently, 
XBLs are seen at angles larger than the RBLs. 
Nevertheless, 
Sambruna et al. (1996) have shown that the typical 
radio-to-X-ray spectrum of an XBL cannot be obtained from the spectrum 
of an RBL simply by changing the viewing angle alone and they suggested that 
other physical parameters, like the intensity of the magnetic fields, 
must be considered.  
Moreover, XBLs and RBLs show significant differences in the slope of their 
X-ray spectra (Comastri, Molendi \& Ghisellini 1995; Urry et al. 1996;  
Padovani \& Giommi 1996; Lamer, Brunner \& Staubert 1996) that are 
difficult to explain in terms of a different viewing angle. 
Giommi \& Padovani (1994) have suggested the existence 
of a unique class of BL Lacs, characterized by a wide range of different 
overall spectral properties.
The X-ray and radio surveys have sampled 
the extreme ends of this distribution thus creating an  
apparent dichotomy: the class of High energy peaked BL Lacs (HBL$\sim$XBL)
and the class of Low energy peaked BL Lacs (LBL$\sim$RBL). 
In this framework, the intermediate objects between HBL and LBL were 
missed in previous X-ray and radio surveys. 
Recently, Fossati et al. (1997) have elaborated further  
the idea proposed by Giommi \& Padovani (1994), suggesting that 
the physical parameter which governs the shape of the spectral distribution 
is the bolometric luminosity.  

The most compelling and straightforward way to test 
these competing hypothesis is to select a new sizable sample of BL Lacs
containing both HBL and LBL, in order to compare {\it 
directly} their properties. 
At present, the only existing statistically complete samples of BL Lacs 
contain only few tens of objects each. 
Since the REX survey reaches radio fluxes 200 times 
fainter than the limiting flux of the 1~Jy sample and X-ray fluxes about 
10 times below the EMSS limits, we expect to find both kinds of BL Lacs as 
well 
as objects with intermediate properties. We have shown the results of 
numerical 
simulations to evaluate the capability of the REX survey to 
select LBLs and HBLs and, as described above, we have found that both kinds 
of objects are expected to be sampled.

\begin{figure}
\plotone{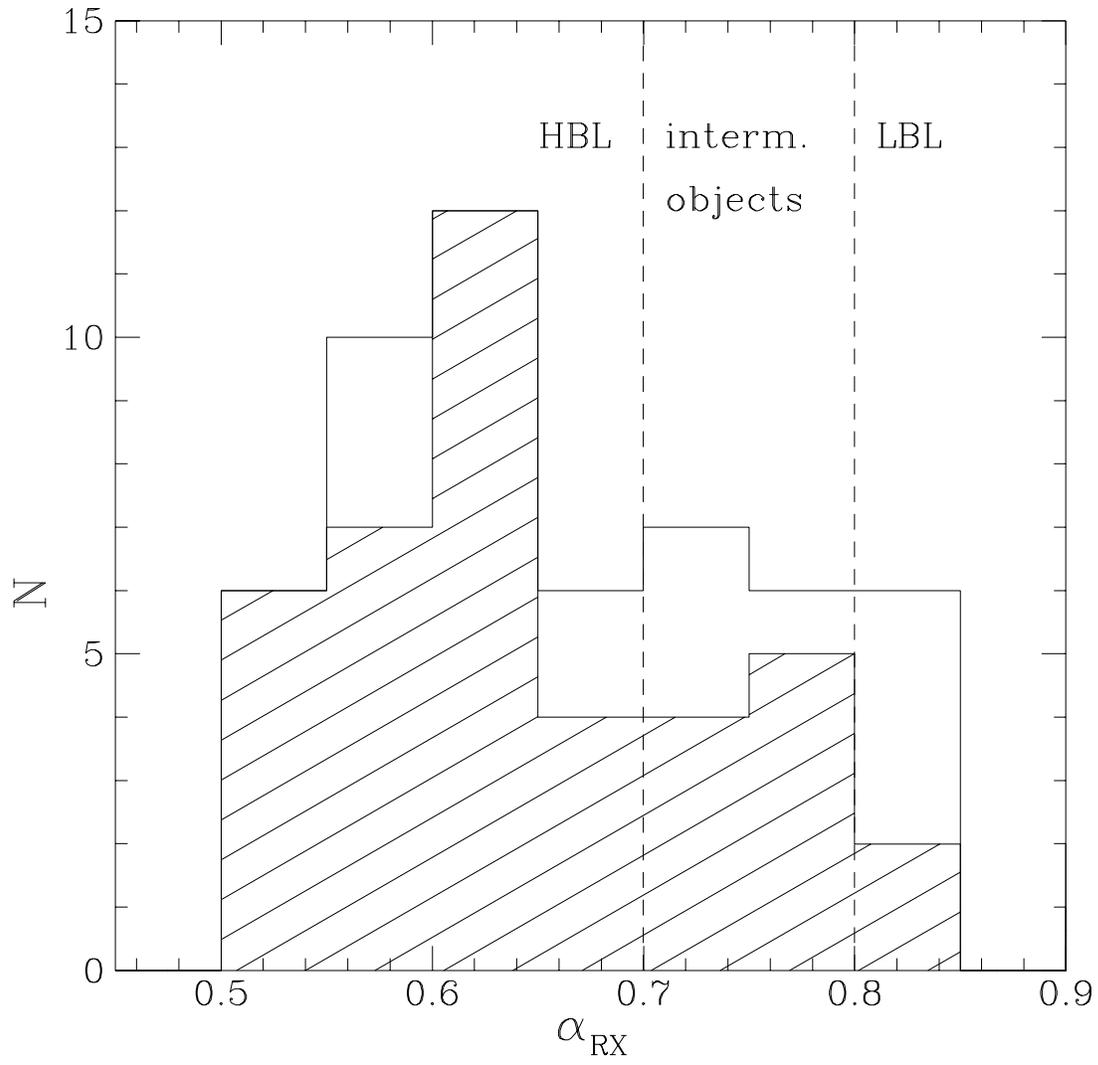} 
\caption{The distribution of the $\alpha_{RX}$ of the BL Lacs in the REX 
sample. The shaded area represents only the ``firm'' BL Lacs, i.e. the 
objects with a Ca~II contrast below 25\% (see text for details).
\label{arxbl}}
\end{figure}

In Figure~\ref{arxbl} we show the $\alpha_{RX}$ 
distribution for the 55 BL Lacs (or BL Lac candidates) discovered 
in the REX survey. The shaded histogram represents only the 
``firm'' BL Lacs. 
This figure does not reveal any bimodality. 
Nearly all the BL Lacs of this sample 
(47/55) fall below the limit on $\alpha_{RX}$ frequently used to 
discriminate between HBL and LBL ($\alpha_{RX}$ = 0.8, e.g. 
Padovani \& Giommi 1996). Apparently, the number of HBL largely exceeds 
that of LBLs, as suggested by the ``viewing angles hypothesis'' or 
the ``bolometric'' scenario proposed by Fossati et al. (1997). 
Moreover, Figure~\ref{arxbl} shows that 
several BL Lacs (13)  are intermediate (0.70$\leq\alpha_{RX}
\leq$ 0.80) between 
the BL Lacs of the EMSS ($\alpha_{RX}\leq$0.70) and that of the 1 Jy sample 
($\alpha_{RX}\geq$0.8). This result
seems to rule out the models proposed by Ghisellini \& Maraschi (1989) or
by Celotti et al. (1993) that predict a sharp separation
between LBLs and HBLs, and to support those models that predict a 
smooth passage between LBLs and HBLs, like those proposed by
Giommi \& Padovani (1994) or by Fossati et al. (1997).
However, the incidence of both the radio and the X-ray limits in the 
selection must be considered before drawing any firm conclusion about 
the relative number of HBL and LBL. The sky coverage of the REX survey, 
in fact, is the combination of the limiting fluxes in the X-ray and in 
the radio band. Other projects, similar to the REX survey, find 
samples of BL Lacs very different in terms of HBL/LBL ratio: for 
example, the DXRBS (Perlman et 
al. 1998) finds a low number of HBL object (16\%) in comparison with 
the intermediate and the LBL objects (84\%). Conversely, the 
RASS-Green Bank sample (RGB, Laurent-Muehleisen et al. 1998) contains 
mostly HBL ($\sim$60\%) and intermediate ($\sim$23\%) objects while only 
the $\sim$17\% is composed by LBLs. Finally, about half of the RC sample 
(Kock et al. 1996) are HBL and about half are intermediate objects. 

These results clearly show the importance of the radio/X-ray limits 
in the sampling of the BL Lac population: the DXRBS is characterized 
by a high ratio between the radio and the X-ray limiting fluxes 
($\alpha_{RX} \sim$0.8--0.9) and, thus, it favours the 
selection of the LBL, while the RGB and the REX surveys are more 
shifted toward lower ratio between radio/X-ray limiting fluxes 
($\alpha_{RX} \sim$0.6--0.7) and, for this reason, they are able to sample 
the HBL population.

A way to take into account these selection effects is 
to compare the spatial densities of the two populations of BL Lacs, 
i.e. the number of objects divided by the volume of universe 
sampled (which must be computed by taking into account the limiting 
fluxes in the two selection bands). Alternatively, it is possible 
to use the competing theoretical models proposed and the 
radio and X-ray sky-coverage of the REX sample to predict the number of BL Lacs
expected in the survey. Then, through a comparison between the predicted 
and the observed number of HBL/LBL it will be possible to draw some 
interesting conclusions about the origin of the BL Lac dichotomy. 
At this stage, however, given the low identification rate of the 
sample, we cannot 
perform any detailed statistical analysis on the ratio HBL/LBL. 

\section{Summary and conclusions}
We have presented and discussed in detail the scientific goals and the 
selection
criteria of the REX Survey, a project optimized for the selection of new, 
large samples
of BL Lacs and Radio Loud AGNs over a large area of sky 
(2183~deg$^2$). Since the existing X-ray catalogs obtained 
from the pointed PSPC observations are not suitable for statistical 
studies we have produced our own X-ray catalogue  from  
the analysis of a well defined subset of all the public PSPC images using a
source detection and
characterization algorithm based on wavelet transforms (Damiani et al. 1997a, 
1997b).
An elaborate positional cross-correlation between this newly compiled 
catalog of X-ray sources and the NVSS (Condon et al. 1998) radio catalog has 
led to 
the definition of about 1450 ``single component'' REX, 
90 REX that show a radio double or triple morphology and about 180 T-REX 
that still need to be correctly classified. Upon
classification of the remaining T-REX, the final number of REXs will be 
close to 1600 (including the objects with a complex radio morphology).
The positional accuracy of the VLA allows us, in the very large majority of 
cases,
to unambiguously identify the optical counterpart of the radio emitting X-ray
source, making feasible a program aimed at the complete identification of
the whole sample. 

The REX sample contains a very high fraction of RL AGNs and BL Lac objects
as indicated by preliminary results of the optical identification program.

The high number of BL Lacs expected in the survey ($\gtrsim$ 200)
will increase significantly our knowledge of their nature and statistical 
properties
currently based on samples of a few tens of objects. Moreover, the possibility
of
comparing directly in the same sample the properties of ``XBL" and ``RBL" will
be
instrumental in testing the existing theoretical emission models.

Finally, unlike other similar projects (e.g. the DXRBS), in the REX survey 
we do not impose any  constraint on the slope of 
the radio spectrum. In this way we will 
be able to compare directly radio galaxies and BL Lacs in the same sample 
and to test the transition between a ``normal'' radio galaxy and a BL Lac
object. The importance of this characteristic is confirmed by the preliminary 
results presented in this paper. We have analyzed a sample of 
newly discovered BL Lac and elliptical galaxies for which we have 
an homogeneous set of optical spectra. Given the high luminosities 
at 1.4~GHz ($\geq$10$^{31}$ erg s$^{-1}$), the elliptical galaxies, that  
do not show any emission line in their optical spectrum, can be 
considered as radio galaxies, probably FR~I. 
In the optical band we have found that the distinction between BL Lacs and 
elliptical galaxies is quite blurred. 
In the X-ray band, both radio galaxies and BL Lacs show 
a significant anti-correlation between the X-ray luminosity and the 
value of the Ca~II contrast (which is an indicator of the fraction between 
the stellar and the nuclear emission in the optical band). The radio galaxies
show higher values of Ca~II contrast ($\sim$50\%) and lower X-ray 
luminosities ($\leq5\times10^{43}$ erg s$^{-1}$) while BL Lacs show lower 
values of
Ca~II contrast ($\leq$40\%) and higher X-ray luminosities 
($\geq5\times$10$^{43}$ erg s$^{-1}$). 
This behavior could be interpreted as due to 
an increasing importance of the nuclear non-thermal components in these 
sources. 
This can be connected either to the intrinsic power of the 
non-thermal nucleus or to the effect of orientation in the framework of the 
beaming model.

\acknowledgments
We thank F. Damiani and S. Sciortino for providing us with their 
source-detection algorithm and for useful discussions and 
C. Ruscica for his contribution in the early stage of this work. 
We also thank the referee Simon Morris for constructive criticisms
and suggestions that significantly improved the quality of the paper. 
J.J.Condon was always very responsive in helping us understand 
all the details of the NVSS. 
This research has made use of the NASA/IPAC extragalactic database (NED), 
which is operated by the JET Propulsion Laboratory, Caltech under contract 
with the National Aeronautics and Space Administration.
We would like to thank the staff of the Laboratory for High Energy Astrophysics 
(LHEA) at NASA/GSFC for their efforts to maintain the ROSAT archive.
This work has received partial financial support from the Italian Space
Agency (ASI), from NASA (NASA grant GO-5402.01-93A, GO-05987.02.94A,
NAG5-2594, NAG5-2914) and from NSF (NSF grant AST95-00515).

\clearpage


\newpage

\newpage

\newpage

\newpage

\newpage

\newpage

\newpage

\newpage

\newpage

\newpage

\newpage

\newpage

\newpage


\begin{deluxetable}{l l l l l}

\small
\tablecaption{Observing setup \label{runs}}
\tablewidth{0pt}
\tablehead{
Telescope/Instrument & Grism name (g/mm) & Slit\tablenotemark{a} & 
Dispersion\tablenotemark{b} & Observing Period}
\startdata
UNAM 2.1m + BC & (300) & 1.6 & 3.9 & 1995 Apr 25--27 \\
UNAM 2.1m + BC & (300) & 1.6 & 3.9 & 1995 Sep 22--24 \\
UH 88''+ WFGS & blue (400) & 1.5 & 4.2 & 1996 Jan 14--15 \\
UH 88''+ WFGS & green (420) & 2.3 & 3.7 & 1996 Aug 7--11 \\
UNAM 2.1m + BC & (300) & 1.6 & 3.9 & 1996 Dec 6--10 \\
ESO 2.2m + EFOSC2 & n.1 (100), n.6 (300) & 1.5 & 13.2, 4.1 & 1996 Dec 11--12 \\
ESO 3.6m + EFOSC1 & B300 (300), R300 (300) & 1.5 & 6.3, 7.5  & 1996 Dec 9--10 \\
UH 88''+ WFGS & blue (400) & 1.5 & 4.2 & 1997 Mar 3--5 \\
UH 88''+ WFGS & blue (400) & 1.5 & 4.2 & 1998 Feb 26 -- Mar 1 \\

\enddata

\tablenotetext{a}{in arcseconds}
\tablenotetext{b}{in \AA/pixel}
\end{deluxetable}

\end{document}